\documentclass[aps,prb,twocolumn,a4paper,eqsecnum,balancelastpage,showpacs,floatfix]{revtex4}
\usepackage{epsfig}
\usepackage{verbatim}
\usepackage{longtable}

\def\dirB{./}

\begin{document}

\title{Relation between driving energy, crack shape and speed in brittle
dynamic fracture}

\author{Andrea Parisi}
\author{Robin C. Ball}
\affiliation{Department of Physics, University of Warwick, Coventry CV4 7AL, United
Kingdom}

\date{\today{}}

\begin{abstract}
We report results on the interrelation between driving force, roughness
exponent, branching and crack speed in a finite element model. We
show that for low applied loadings the crack speed reaches the values
measured in the experiments, and the crack surface roughness is compatible
with logarithmic scaling. At higher loadings, the crack speed increases,
and the crack roughness exponent approaches the value measured at
short length scales in experiments. In the case of high anisotropy,
the crack speed is fully compatible with the values measured in experiments
on anisotropic materials, and we are able to interpret explicitly the
results in terms of the efficiency function introduced by us in our
previous work {[}A.~Parisi and R.~C.~Ball, \emph{Phys.~Rev.~B}
\textbf{66}(16) 165432 (2002){]}. The mechanism which leads to the
decrease of crack speed and the appearence of the logarithmic scaling
is {\it attempted} branching, whilst power law roughness develops when
branches succeed in growing to macroscopic size.
\end{abstract}


\maketitle

\section{Introduction\label{sec:introduction}}

There are two questions in the field of fracture mechanics to which
intense research has been devoted in recent years. The first concerns
the terminal speed predicted by the continuum theory \cite{F-90}
which does not match the maximum speed measured in 
experiments.\cite{KOS-74,FGMS-91,FGMS-92}
Terminal crack speeds are usually found to vary in a range between
about $90\,\%$ of the Rayleigh speed in anisotropic 
materials,\cite{F-71,HB-66,WK-94}
down to values as low as $33\,\%$ for more isotropic
materials.\cite{KOS-74,FGMS-91,GFMMS-93}
The Rayleigh surface wave speed $v_R$ is 
the terminal crack speed expected in the continuum theory;\cite{F-90,FM-99} beyond it the continuum solutions show compressive sign to what should be the crack opening stress component ahead of the crack tip.\footnote{
This is displayed explicitly for Yoffe's crack solutions\cite{Y-51} in Ref.~\onlinecite{F-90} and can also be inferred from the results for climbing edge dislocations in Ref.~\onlinecite{WW-80}.} The experimental maximum speed is usually accompanied by tip branching
which is also not well explained. What controls the instability which
leads to branching? What sets the terminal crack speed? Both the continuum
theory and computer investigations have suggested that the answer
to these questions lies in the mechanism through which energy is dissipated
at the crack tip.\cite{FM-99}  

The second question involves the roughness exponent governing the
self-affine scaling of height fluctuations of fracture surfaces. Recent
measurements \cite{DNBC-97} have shown how this quantity has a universal
behaviour in that it takes a value of $0.5$ at {}``short length
scales'' (in experiments this usually corresponds to nanometer scales)
and a value of $0.75$ at {}``large scales'', the two regimes being
separated by a material dependent crossover length. 
In addition, a logarithmic scaling of fracture surfaces has been 
theoretically\cite{BL-95,REF-97} and experimentally\cite{LB-95} found 
in the limit of quasi-static crack advance in brittle materials.
Despite numerous
attempts to describe the universal character of such behaviour, the
question of what controls the value of the roughness exponent and
why fractures tend to grow rough surfaces is still open.

The two problems have repeatedly appeared to be connected.
Experiments performed on polymethilmethacrylate (PMMA) have shown
that beyond a critical crack speed, the crack tip starts to oscillate 
leading to the formation of structures on the crack surface and to 
a departure of the crack speed from that expected from the continuum
theory.\cite{FGMS-91,FGMS-92}  
The same phenomenology was observed in simulations, where departure 
from steady state propagation for cracks exceeding a threshold 
speed was observed, with zig-zag motion and formation of 
microstructures.\cite{ABRR-94}
Molecular dynamics simulations of crystalline silicon showed that 
cracks can dissipate 
large amounts of energy, up to seven times the energy needed to create
a smooth surface as estimated in the framework of the continuum
theory, the suggestion being that this energy goes into 
lattice oscillations.\cite{HM-98b}  

The idea that the energy available does not all go into fracture
work is not new.
Analytical studies of planar crack advance in a lattice by 
Slepyan had already shown that the presence of
the lattice leads to an important excess of energy being radiated 
from a crack at both low and high speeds,\cite{S-81} and that crack 
propagation at low speeds is unstable\cite{S-83}.  
More recently, Marder and Liu \cite{ML-93} have studied a lattice model 
for fractures, concluding that it is lattice oscillations 
which limit the range of possible crack speeds.\cite{ML-93,M-99}

In a recent paper\cite{PB-02} we have shown how such energy radiation
at the crack tip due to phonon emission is the crucial mechanism 
for crack propagation.  
The intensity of the radiated energy is 
a function of the crack speed, ruling out stable crack growth at low speeds. 
The dependence of the crack speed is in agreement with the analytical
results of Slepyan.\cite{S-81,S-83}  The instability at low crack speeds
also corresponds to what suggested in Ref.~\onlinecite{S-83} and to the
results obtained by Marder and Liu.\cite{ML-93,M-99}
Our results suggest a way to relate the intensity of the radiated 
energy directly to the phonon band structure.
The relation between this approach and the
results of Marder {\em et al.}\cite{M-99,HM-99,FM-99} is discussed in
Ref.~\onlinecite{PB-02}.  The mechanism proposed has been
investigated in the limiting case of planar cracks, but the analysis
becomes more difficult in presence of branching.  However there is clear simulation evidence in two dimensions that crack branching is sensitive to the lengthscale at which the continuum description breaks down.\cite{FNR-01}

The mechanism that leads to branching has not yet been fully unravelled.
The first hint to the understanding of such phenomenon came from
Yoffe\cite{Y-51} who showed that beyond a critical crack speed, 
the hoop stress has a maximum at a definite angle
with respect to the direction of crack propagation.  This has attracted considerable discussion as experiments have shown 
branching at crack speeds which differ from the prediction 
of Yoffe.\cite{CK-84,FGMS-92,GFMMS-93} Simulations and experiments however seem to 
agree that the mechanism of branching could be connected to the terminal 
crack speed.\cite{FGMS-92,XN-94,MERZ-00}  

In this work we will show that in our simulations the main mechanism that 
limits the crack speed for a free running crack is attemped branching.
Cracks constrained on a plane reach high speeds compatible with the speeds
measured in highly anisotropic materials, which we are able to explain 
in terms of the efficiency function introduced in our previous work.\cite{PB-02} 
We will also show that branching is responsible for the roughness exponent 
of the fracture surfaces: attempted branching roughens the crack surface 
with a logarithmic scaling first and, when macroscopic branches develop, 
with a roughness exponent close to the value measured 
at short length scales.  At the same time, the crack speed is drastically reduced
due to the attempted branching mechanism to values comparable with those measured
in experiments, and gradually rejoins the efficiency description for high loadings.

The model used for these simulations has been extensively described in our
previous work,\cite{PB-02} and its distinctive features are reviewed in 
section \ref{sec:model}.  In section 
\ref{sec:disorder} we discuss the origins and effect of both disorder and
anisotropy, and show how to implement disorder in the model and 
control the driving energy.  Results on the crack speed for both planar 
and non-planar cracks are described in section \ref{sec:crack_speed}, whilst
the scaling of crack surface roughness for different driving regimes is discussed 
in section \ref{sec:crack_roughness}.  An attempt to simulate disconnected
fractures is reported in section \ref{sec:disconnected} and finally we 
draw conclusions in section \ref{sec:conclusions}.

\begin{figure}
\centerline{
\epsfig{figure=\dirB 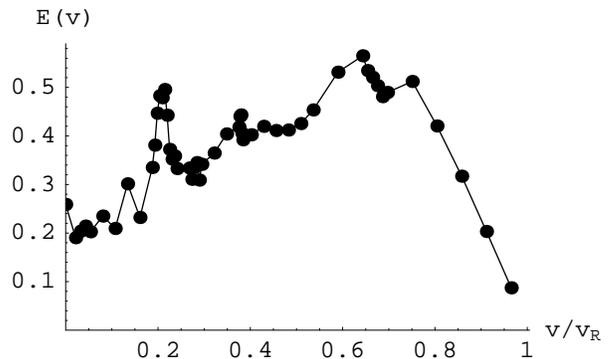, width=8cm}
}
\caption{\label{fig:efficiency} The efficiency for planar crack propagation in our fcc lattice model. The overall
increase of $E(v)$ with the crack speed still remains to be quantitatively
understood, but the drop of the efficiency at the Rayleigh speed as
well as the fine structure are well understood in terms of vibrational resonances.\centerline{ }}
\end{figure}

\section{The model\label{sec:model}}

The model used in these simulations is an application of the finite
element scheme and has been fully described in Ref.~\onlinecite{PB-02}.
The elastodynamic description is obtained by discretizing space on an 
fcc lattice and connecting neighbouring lattice points with (non-filling) 
tetrahedral elements.  We then solve the Euler-Lagrange equations obtained 
by using the discretized form of the Lagrangian of continuum elasticity:
\begin{displaymath}
L = \sum_v \frac{m \dot{\bf u}_v^2}{2} - \sum_t \frac{1}{2} \Omega' (\nabla 
    {\bf u})_t : \stackrel{\leftrightarrow}{\sigma}_t 
\end{displaymath}

\noindent
where ${\bf u}_v$ is the displacement field at site $v$ and 
$\stackrel{\leftrightarrow}{\sigma}_t$ is the stress tensor at 
element $t$, the index $v$ spanning all lattice points and the index $t$
spanning all tetrahedral elements.
The stress tensor is related to the displacement field by the standard tensorial 
relation of continuum elasticity 
$
\stackrel{\leftrightarrow}{\sigma} = \lambda \textrm{Tr(} \nabla {\bf u}
         \textrm{)}\,\openone + \mu [ \nabla {\bf  u} + (\nabla {\bf u})^T ]
$
and $\Omega'$ is a volume element related to the discretization scheme used. 
No mechanism of dissipation is active, but due to the discreteness introduced
waves are radiated from the crack tip with an intensity which is non linear
in the crack speed, revealing a selection rule for the crack speed itself.

Rupture is simulated by irreversibly setting the elastic constants 
$\lambda$ and $\mu$ to zero in any tetrahedral element where
the elastic energy stored within becomes greater than a pre-determined 
fracture energy.  This is not the only possibility for a rupture
criterion: quite recently, Heizler, Kessler and Levine\cite{HKL-02,KL-01} 
have studied the consequence of having a set of continuous force laws
as opposed to the usual piecewise discontinuous
force laws we use.  Their results suggest 
that the nature of the force law could change 
the stability limit for high speed crack motion and, 
at least in mode-I fractures, for low crack motion.  
Using finite elements, a continuous force law was used by
Johnson\cite{J-92} to study the extension of the process
region for moving cracks.
Needleman implemented crack rupture in a finite element model 
using surface decohesion.\cite{N-87}  The model was used both 
in the study of crack growth in brittle 
solids\cite{XN-94} and in the study of
interfacial crack growth.\cite{XN-95a,XN-95b}
It was found that a crack constrained to a plane can approach 
the Rayleigh speed, whilst when unconstrained, crack branching and reduction of the terminal
crack speed were found.  A similar behaviour is found in
our simulations.

Our model has been used in our previous work in the simplified case
of two dimensions, to simulate planar cracks running at fixed speed.
In the present article, we present results of full three-dimensional
simulations, in which both the condition of fixed crack speed and
planarity are released, and the crack advance is only controlled by
a Griffith criterion. In the continuum description, the Griffith criterion
states that a crack will advance only if the macroscopic energy delivered
to the crack tip $G_{M}(v,t)$ exceeds the fracture work $2\gamma_{0}$
necessary to create new surface: \[
G_{M}(v,t)\geq2\gamma_{0}\]

\noindent If $G_{M}(v,t)>2\gamma_{0}$, the excess of energy is usually
considered the source of kinetic energy for crack advance, so if the
loading is just sufficient to have the crack propagate, the crack
should advance quasi-statically. Both simulations and experiments
show that this is not the case: cracks do accelerate rapidly towards
a limiting crack speed which depends on the 
material.\cite{AH-76,SDPM-88,FGMS-92}

The solution of the puzzle is the presence of an alternative mechanism
of energy dissipation due to the discreteness of matter at the microscopic
level, a mechanism which is not included in the continuum elastodynamic
description. Due to the discreteness, the macroscopic energy release
rate $G_{M}(v,t)$ can be equated to the sum of two microscopic contributions:
\[
G_{M}(v,t)=G_{\textrm{\scriptsize{br}}}(v,t)+G_{\textrm{\scriptsize{ph}}}(v,t)\]

\noindent 
where $G_{\textrm{\scriptsize{br}}}(v,t)$ and $G_{\textrm{\scriptsize{ph}}}(v,t)$
are respectively the breakage energy release rate, which is the portion
of the available energy going into fracture work, and the phonon energy
release rate which is the portion of the available energy radiated
as phonons. The breakage energy release rate can be expressed as:
\begin{equation}
G_{\textrm{\scriptsize{br}}}(v,t)=E(v)G_{M}(v,t)
\label{eq:efficiency}
\end{equation}

\noindent 
by introducing the efficiency $E(v)$.  In the case of a strip geometry
with fixed displacements at the top and bottom boundaries, the 
macroscopic energy release rate is a constant independent of the
crack speed:\cite{F-90} $G_{M}(v,t)=G^{\infty}_M$.  Hence, for this
special case the efficiency becomes the sole source of velocity
dependence, thus separating the effect of local discreteness from 
the effect of the macroscopic external loading.  Moreover, the 
efficiency $E(v)$
has been shown to depend only on the lattice geometry and crack
speed, being local to the crack tip and independent of the macroscopic 
dynamical history.\cite{PB-02}

The speed dependence of the efficiency for our fcc lattice model
is shown in fig.~\ref{fig:efficiency}. Although the overall increase
of $E(v)$ with the crack speed still remains to be quantitatively
understood, the drop of the efficiency at the Rayleigh speed as well
as the fine structure are well understood and correspond to resonant
emission, when emitted waves have a group velocity matching the crack
speed itself. In particular, we expect the drop at the Rayleigh speed
to be a feature common to all materials, as that resonance arises
in the continuum limit.
The efficiency $E(v)$ for energy transfer into bond breakage is well
below unity even for zero speed, as can clearly be seen in figure
\ref{fig:efficiency}. This is because when a lattice element (tetrahedron)
breaks at the crack tip, others around recoil dynamically and that
recoil energy is ultimately radiated as sound waves.


A crack will only advance if the energy available at the crack tip
is sufficient to create new surface.  This translates the Griffith 
criterion in presence of discreteness into a condition of 
steady crack growth given by:
\begin{equation}
G_{\textrm{\scriptsize{br}}}(v,t) = 2 \gamma_0,
\label{eq:Griffith-discr}
\end{equation}

\noindent 
the difference from the macroscopic energy release rate
being converted into phonons.
In the case of a strip geometry with fixed displacements at the top 
and bottom boundaries, the threshold to initiate crack advance
is given by $\displaystyle G_{M}^{\infty}=2\gamma_{0}/E(0)$ and
therefore if the loading is maintained only cracks with speed such
that $E(v) = E(0)$ can propagate, leading to $v \simeq 0.88\,v_{R}$ from
the data of figure \ref{fig:efficiency}. We have further argued in
Ref.~\onlinecite{PB-02} that only the more limited regions where 
${dE}/{dv}<0$ should be sustainable.

One can of course load a sample above the quasistatic threshold, with
\begin{displaymath}
G_{M}^{\infty}=\epsilon\frac{2\gamma_{0}}{E(0)}
\end{displaymath}
where $\epsilon>1$. In this case the condition for sustained crack
propagation becomes 
\begin{equation}
E(v) = E(0) / \epsilon
\label{eq:eff-equil}
\end{equation}
and by reference
to fig.~\ref{fig:efficiency} we see that the higher loading can sustain
higher crack speeds, as might be expected.

As the efficiency only depends on the crack speed, by using 
eq.~(\ref{eq:efficiency}) we can calculate $G_{\textrm{\scriptsize{br}}}(v,t)$
for any macroscopic loading for which $G_M(v,t)$ is known, 
and use it in eq.~(\ref{eq:Griffith-discr}) to get the corresponding 
allowed crack speeds.

%
\section{The role of disorder and anisotropy\label{sec:disorder}}

Discretization naturally introduces preferred directions in space
and therefore simulations are characterized by some level of anisotropy.
In the absence of disorder, a square (cubic) lattice acts as a planar
guide for the advancing crack. This phenomenon, known as \emph{lattice
trapping}, was predicted by Thomson, Hsieh and Rama\cite{THR-71} and
described by Holland and Marder\cite{HM-98a} who observed it in
molecular dynamics simulations of silica samples. Experiments also
show that it is possible to obtain atomically flat fracture surfaces
in real crystalline materials by using sufficiently small and homogeneous
loadings.\cite{HHMS-99}

The majority of the experiments on cracks produce non-planar,
rough and branched cracks. The departure from the planar geometry
forced by the lattice trapping is due to two different contributions.
First, the magnitude of the applied loading which influences the energy
delivered to the crack tip: the higher the value, the larger the possibility
for the crack to open out of plane branches due to the increasing
transversal stresses. Second, disorder in the material (equivalently
disorder in the breakage rule) can drive the crack on non-planar paths.

Disorder is naturally present in all materials and comes from a variety
of different sources. Atomic vacancies, inclusions, dislocations 
and grain boundaries are all sources of disorder
able to influence the macroscopic response. Disorder strongly reduces
the effects of anisotropy by increasing the probability of deviations
from planarity. This is the main reason  why a high level of
disorder was included in the simulations presented in this article.
The easiest way to introduce it (and the way followed by us) is
to introduce a locally variable fracture energy $\gamma({\textbf{x}})$
which varies according to some well defined distribution: although
very simple, the uniform distribution accomplishes this task extremely
well. If $\gamma_{0}$ is the fracture energy in the absence of disorder,
local fracture energies can be extracted by a uniform distribution
centered on $\gamma_{0}$ and ranging between $0$ and $2\gamma_{0}$;
this ensures that the mean fracture energy corresponds to the fracture
energy in the absence of disorder. 

In the presence of disorder even a static planar test crack has different
tetrahedra along the crack tip becoming breakable at different loadings,
whereas without disorder they all became breakable at the same loading
${2\gamma_{0}}/{E(0)}$. With disorder we defined the reference
loading (corresponding to $\epsilon=1$) to be the value of $G_{M}^{\infty}$ at
which 50\% of crack tip tetrahedra in a planar static test crack are
not breakable.

In practice modestly lower loadings (e.g. $\epsilon=0.7$) can still
lead to crack propagation, as breakage of vulnerable tetrahedra along
the crack edge leads to stress concentration at and around more resistant
tetrahedra. Higher values of $\epsilon$ lead to more heavily damaged
samples.

%
\section{Free-running fracture simulations in three dimensions\label{sec:crack_speed}}

We now focus on cracks when the constraints of fixed crack speed and
shape are released. A fixed displacement is applied to both top and bottom faces, and a starting notch is prepared on the front
face. The starting notch is long enough to start the simulations in
the long crack limit. Periodic boundary conditions are applied to
the side faces. The front and back faces are left stress free.

Each tetrahedron in the sample was given a pre-assigned fracture energy
drawn from a broad uniform distribution centered on $\gamma_{0}$ as described 
in the previous section. At each
timestep the energy of each tetrahedron is evaluated, and those in
which this exceeds their fracture energy are broken by setting the
two elastic constants $\lambda=\mu=0$. 
The presence of disorder leads to breakage of isolated tetrahedra 
as soon as the simulation starts.  A condition of connected fracture
is imposed by allowing only neighbours of already broken tetrahedra
to break.  This simplifies the track of the crack tip in order to measure
its speed.  The condition of connected fracture prevents the formation of
precracks in front of the crack tip.  In practice, if this condition is 
released there is essentially no difference in the results for moderated 
loadings.  We will see in section \ref{sec:disconnected} that for high 
loadings this can lead to a different morphology for the averall 
fracture process.

\begingroup
\squeezetable
\begin{table}
\begin{ruledtabular}
\begin{tabular}{c c c}
\multicolumn{3}{c}{(a) Planar cracks with disorder}\\
\hline
$\epsilon$ & Histogram meas. & Average tip meas.\\
\hline
$0.7$ & $0.807 \pm 0.010$ & $0.810 \pm 0.018$ \\
$1.0$ & $0.876 \pm 0.008$ & $0.866 \pm 0.017$ \\
$2.0$ & $0.933 \pm 0.003$ & $0.922 \pm 0.011$ \\
$3.0$ & $0.945 \pm 0.009$ & $0.942 \pm 0.016$ \\
$4.0$ & $0.951 \pm 0.013$ & $0.951 \pm 0.033$ \\
$5.0$ & $0.988 \pm 0.013$ & $0.995 \pm 0.055$ \\
\hline
  & & \\
\multicolumn{3}{c}{(b) Non-planar cracks: $300 \times 60 \times 60$} \\
\hline
$\epsilon$ & Histogram meas. & Average tip meas.\\
\hline
$0.7$ & $0.552 \pm 0.029$ & $0.546 \pm 0.004$ \\
$1.0$ & $0.724 \pm 0.002$ & $0.720 \pm 0.003$ \\
$2.0$ & $0.824 \pm 0.001$ & $0.798 \pm 0.013$ \\
$3.0$ & $0.912 \pm 0.008$ & $0.839 \pm 0.005$ \\
$4.0$ & $0.982 \pm 0.005$ & $0.908 \pm 0.004$ \\
$5.0$ & $1.079 \pm 0.047$ & $0.940 \pm 0.018$ \\
\hline
  & & \\
\multicolumn{3}{c}{(c) Non-planar cracks: $500 \times 120 \times 120$} \\
\hline
$\epsilon$ & Histogram meas. & Average tip meas.\\
\hline
$0.7$ & $0.545 \pm 0.004$ & $0.549 \pm 0.004$ \\
$1.0$ & $0.700 \pm 0.001$ & $0.693 \pm 0.003$ \\
$2.0$ & $0.782 \pm 0.003$ & $0.757 \pm 0.011$ \\
$3.0$ & $0.833 \pm 0.008$ & $0.792 \pm 0.001$ \\
$4.0$ & $0.876 \pm 0.009$ & $0.811 \pm 0.002$ \\
$5.0$ & $0.905 \pm 0.008$ & $0.832 \pm 0.006$ \\
\end{tabular}
\end{ruledtabular}
\caption{\label{tab:crack-speeds}
Measurements of the crack speed $v/v_{R}$ for different driving energies,
in different types of simulations: (a) cracks forced to be planar,
with the addition of disorder in the threshold energy; (b) free cracks
with disorder for samples $300\times60\times60$ tetrahedra wide;
(c) free cracks with disorder for samples $500\times120\times120$
tetrahedra wide. In all tables, the first column is the driving factor
$\epsilon$, the second and third columns are the crack speed in units
of the transverse wave speed as measured by the histogram method (second
column) and the average tip position method (third column). For all
simulations, $v_{R}\simeq0.933\,\, v_{t}$.}
\end{table}
\endgroup

\begin{figure}
\centerline{
\epsfig{figure=\dirB 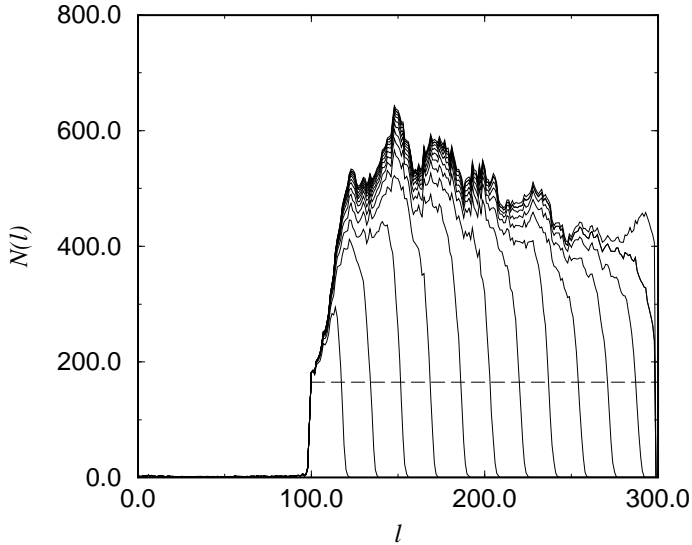, width=10cm}
}
\caption{The histogram
method for measuring the crack speed. The histogram represents the
number of broken tetrahedra $N(l)$ at distance $l$ from the front
side of the sample. As time advances, the histogram grows: we
can build a set of histograms corresponding to a set of intermediate
positions of the crack. The approximate position of the crack front
is then obtained as the intercept with the dashed threshold.\label{fig:crackspeed-histo} }
\end{figure}

\begin{figure}[b]
\centerline{
\epsfig{figure=\dirB 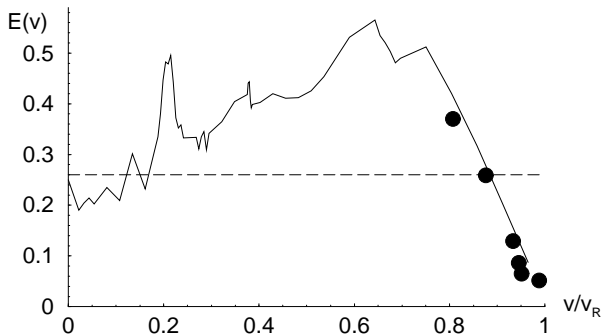, width=8cm}
}
\caption{\label{fig:crackspeed-energy}Comparison between the results 
for free-running planar cracks (filled circles) and the efficiency 
description (continuous line).
Higher values of the driving energy $\epsilon$ correspond to lower
values of $E(v)$ as described by eq.~(\ref{eq:eff-equil}).
The figure shows how this also corresponds to higher crack speeds.
The dashed line refers to the special case of $\epsilon=1$.
}
\end{figure}

\begin{figure*}
\centerline{
   \epsfig{figure=\dirB 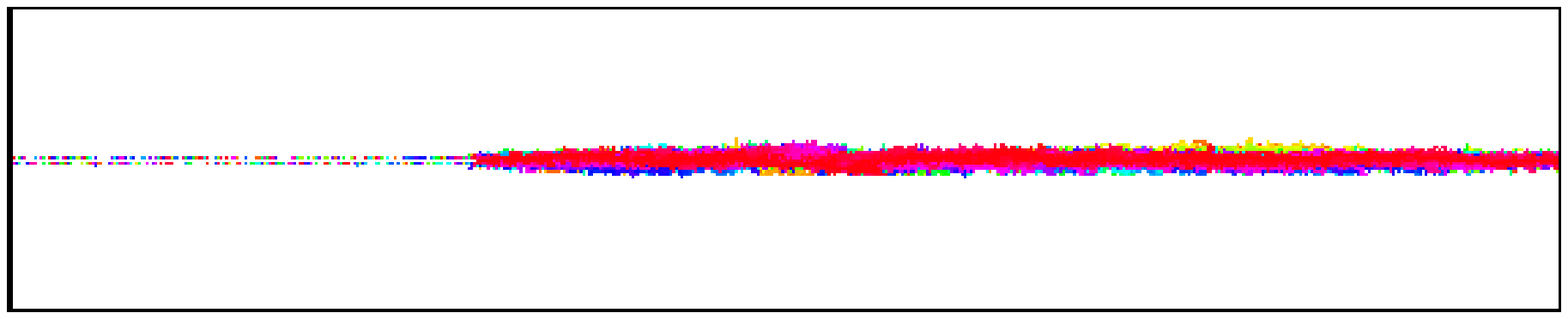, width=13cm}
}
\medskip
\centerline{
   \epsfig{figure=\dirB 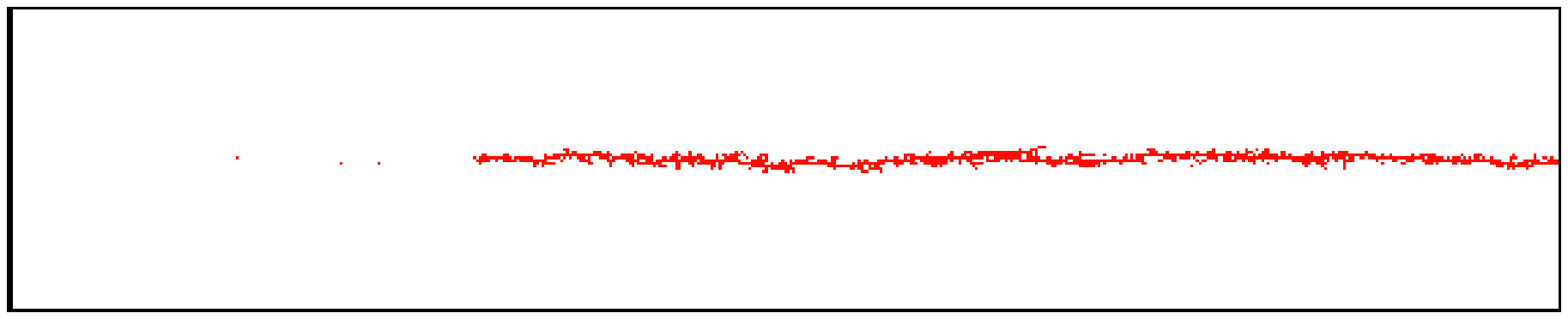, width=13cm}
}
\caption{(Color online) (Top) Crack in a sample $300\times60\times60$ tetrahedra wide 
(projection), for $\epsilon=1$. Shading (and color online) 
refers to the depth in the third dimension.
The crack appears {}``fat'',
thicker than one layer of tetrahedra. A single-layer wide section
of the sample (bottom) shows that the crack itself is driven
slightly off planar. Microbranching is visible, suggesting attempted
branching as the mechanism which leads to thicker cracks and slows
the crack speed. \label{fig:loweps} }
\end{figure*}

\begin{figure}[b]
\centerline{
   \epsfig{figure=\dirB 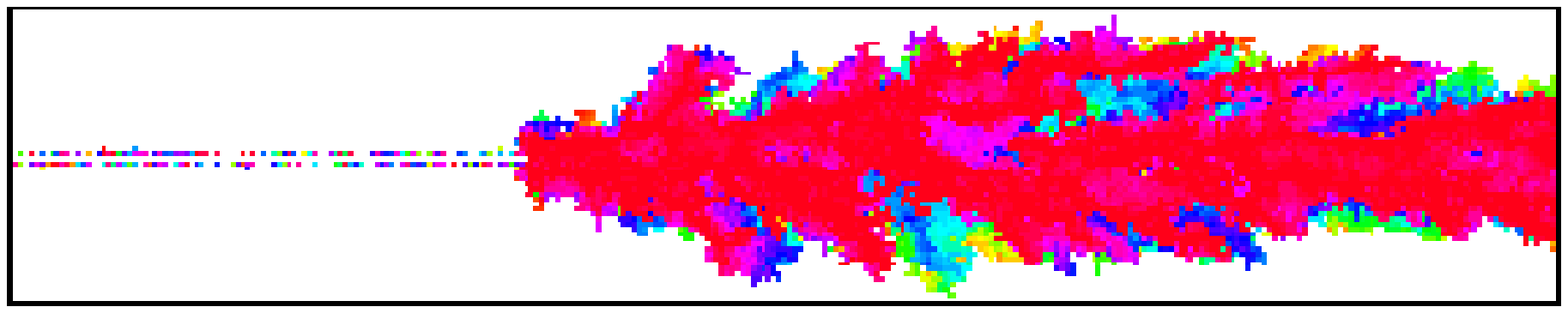, width=8cm}
}
\medskip
\centerline{
   \epsfig{figure=\dirB 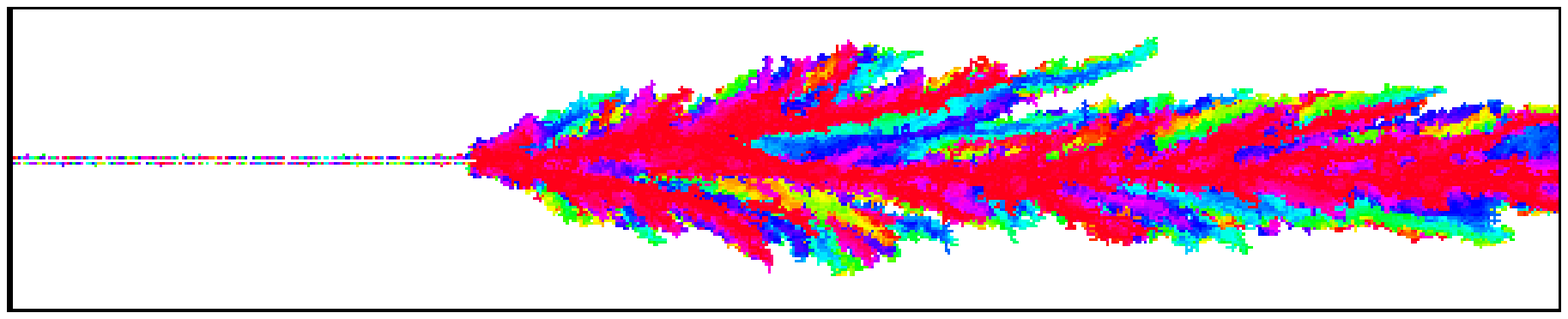, width=8cm}
}
\medskip
\caption{\label{fig:sample-eps5}(Color online) (Top) Crack in a sample 
$300 \times 60 \times 60$ tetrahedra wide (projection),
$\epsilon=5$: branches reach the sample's boundaries. (Bottom) Crack in
a larger sample, $500\times120\times120$ tetrahedra wide (projection), for $\epsilon=5$.
In this case branches do not reach the sample boundaries. Shading 
(and color online) refers to the depth in the third dimension.
}
\end{figure}

Each simulation starts from the sample relaxed to its configuration
of minimum energy, and continues until the sample is broken into two
halves. We wait until the number of broken tetrahedra per timestep
is reduced to a negligible fraction of its peak and then the sample
is divided into two halves according to the sign of the vertical displacement
of the sites. In some cases the sample remains connected by a few
isolated tetrahedra, but as these will be in a state of anomalously
high strain, the displacement of their sites will reflect the displacement
of the portion of sample to which they are attached.

That the surface found corresponds to the fracture surface has been
checked using a different method based on percolation. The sample
can be described as a collection of boxes (the tetrahedra) connected
on a cubic lattice. If we choose one tetrahedron on the bottom face
we can imagine injecting it with some coloured liquid: the liquid
will spread within the sample and reach the top face. Broken tetrahedra
can act as blocking boxes for the liquid, so that the liquid will
reach the top face unless a complete fracture surface dividing the
sample into two halves is retrieved. Complete failure was assured
by slowly increasing the applied displacement until percolation between
top and bottom was lost, requiring significant extra simulation time.
Comparing the two methods for a set of $10$ simulations a difference
of less then $0.05\%$ for the set of broken bonds with respect to
the whole sample was found, and similarly a difference in the final
detected surface of less than $0.9\%$. %

\begin{figure*}
\centerline{
\epsfig{figure=\dirB 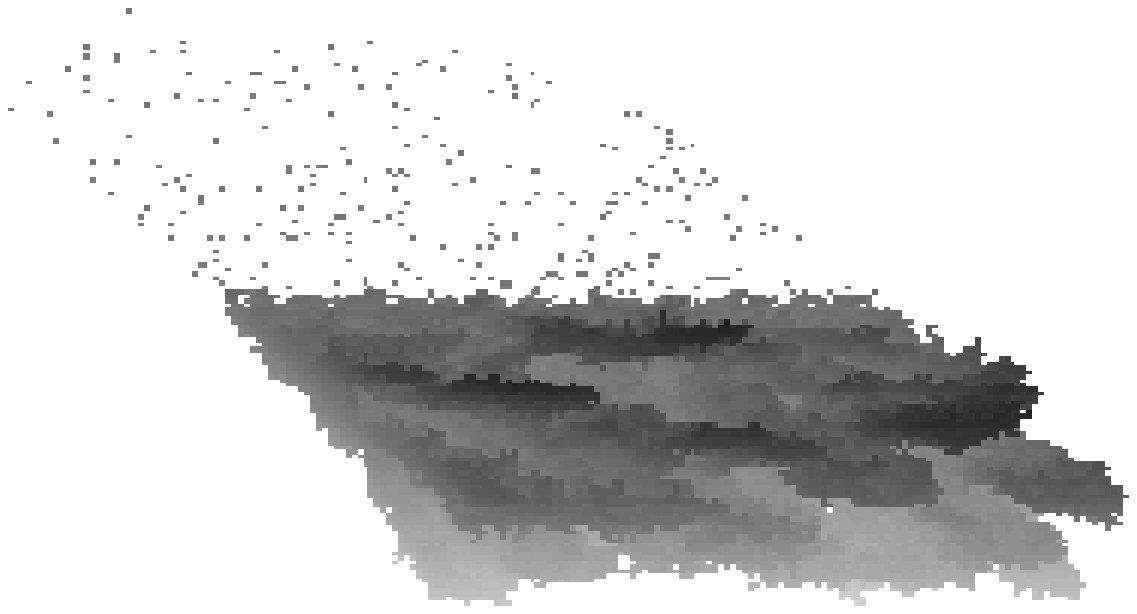, width=6cm}
\epsfig{figure=\dirB 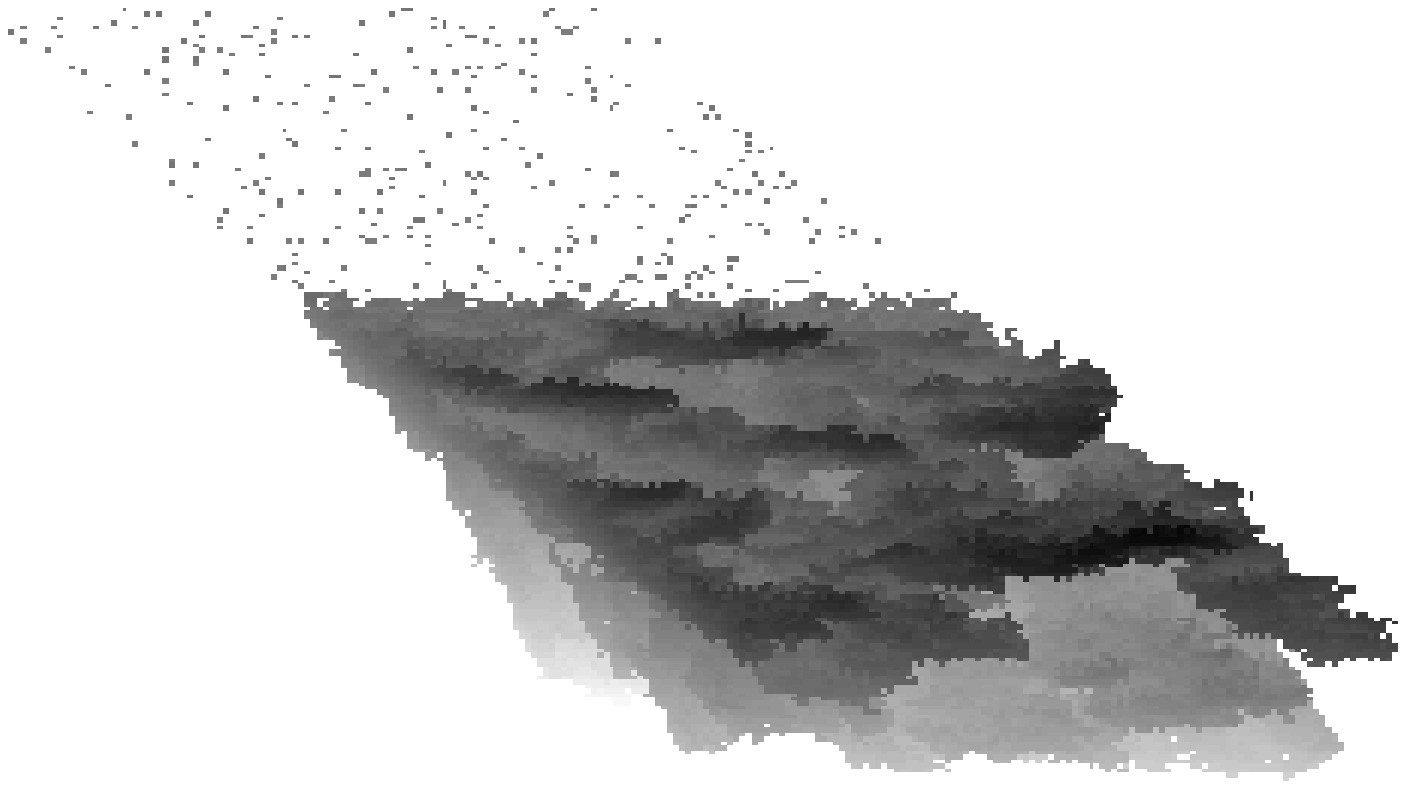, width=6cm}
}
\centerline{
\epsfig{figure=\dirB 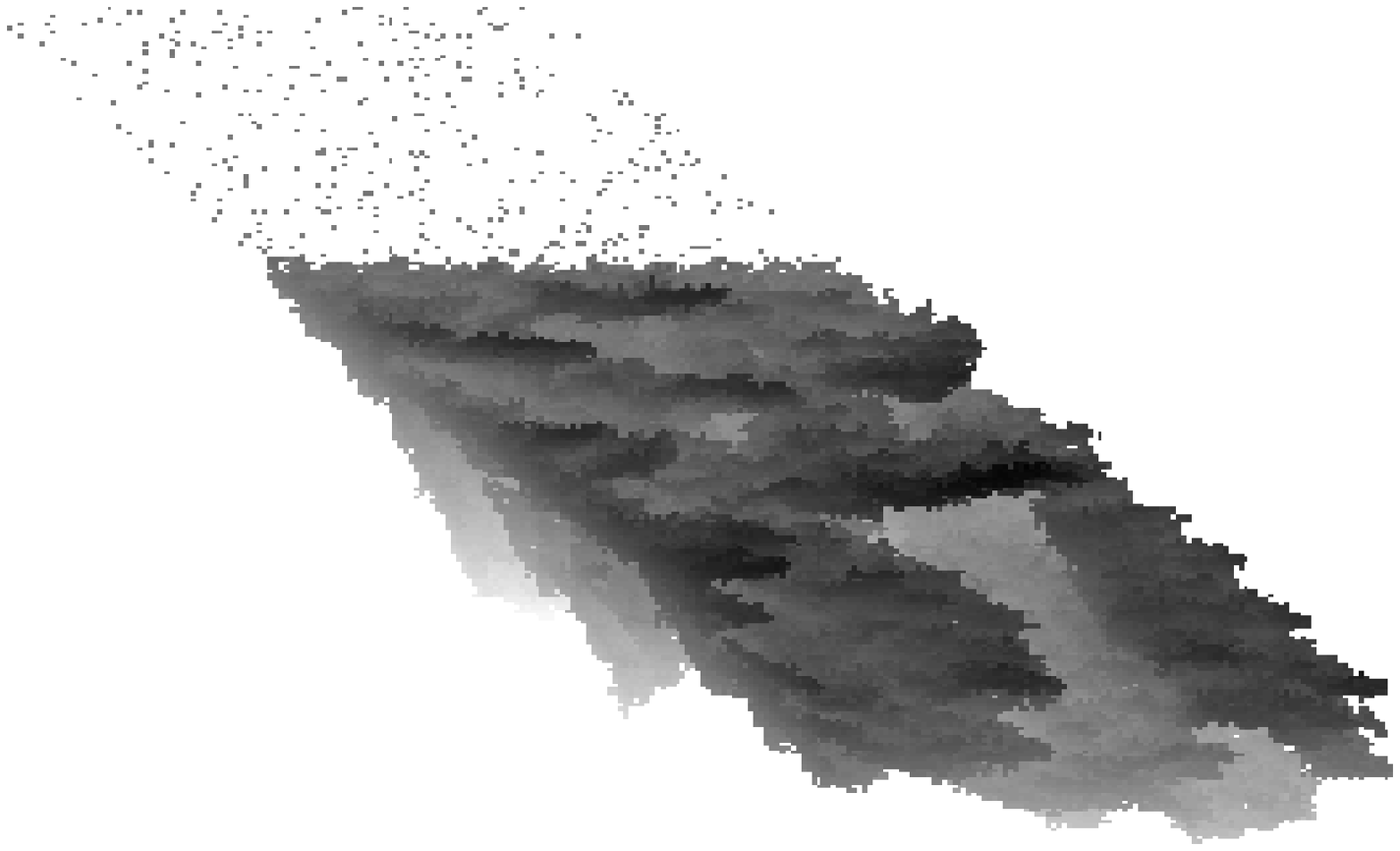, width=6cm}
\epsfig{figure=\dirB 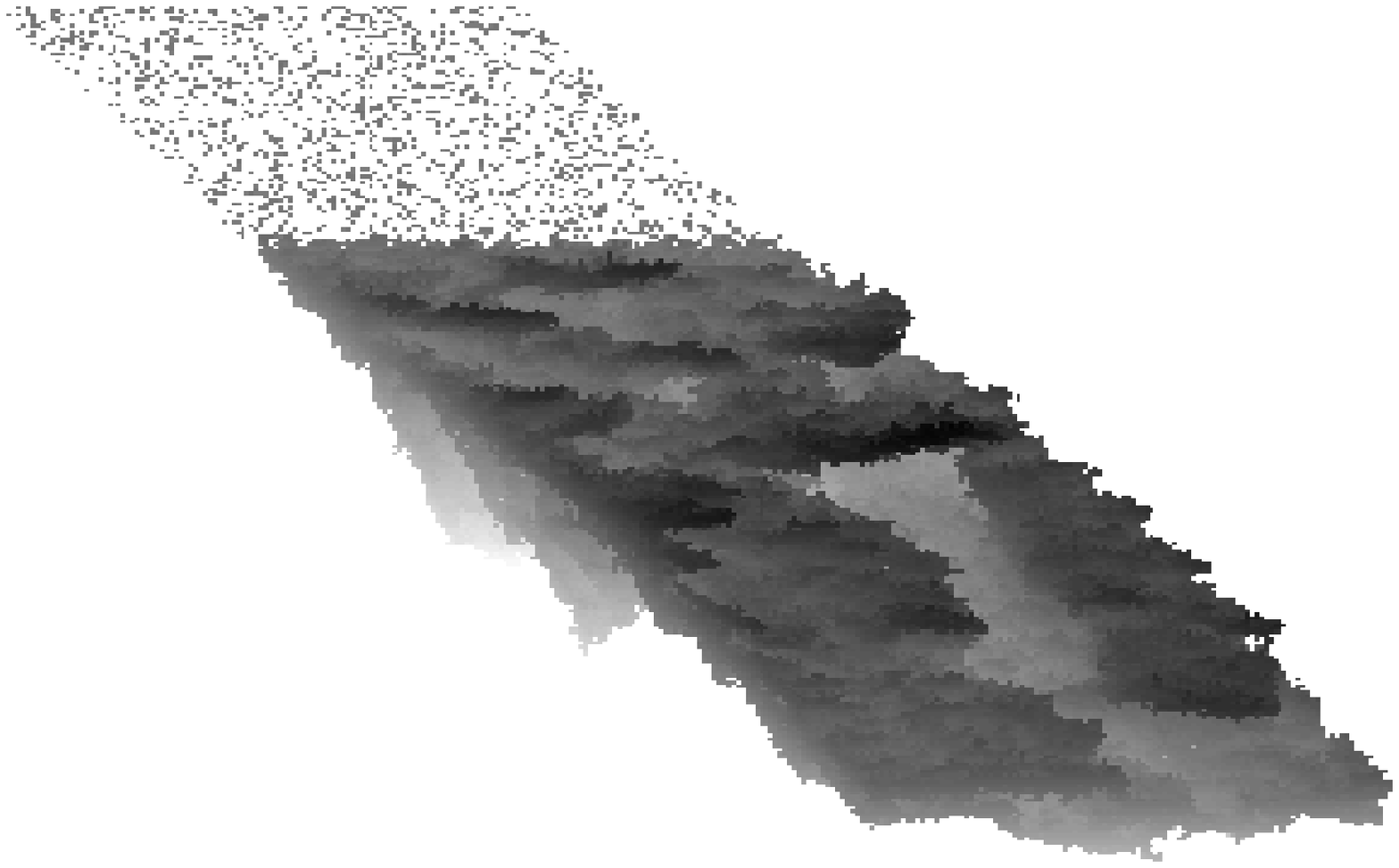, width=6cm}
}
\caption{\label{fig:sequence}Sequence of frames for a typical crack growing 
in a sample 
$500\times120\times120$ tetrahedra wide. The gray shading refers to 
the height with respect to the crack notch plane.
In this case $\epsilon=3$.}
\end{figure*}

\begin{figure}[b]
\centerline{
\epsfig{figure=\dirB 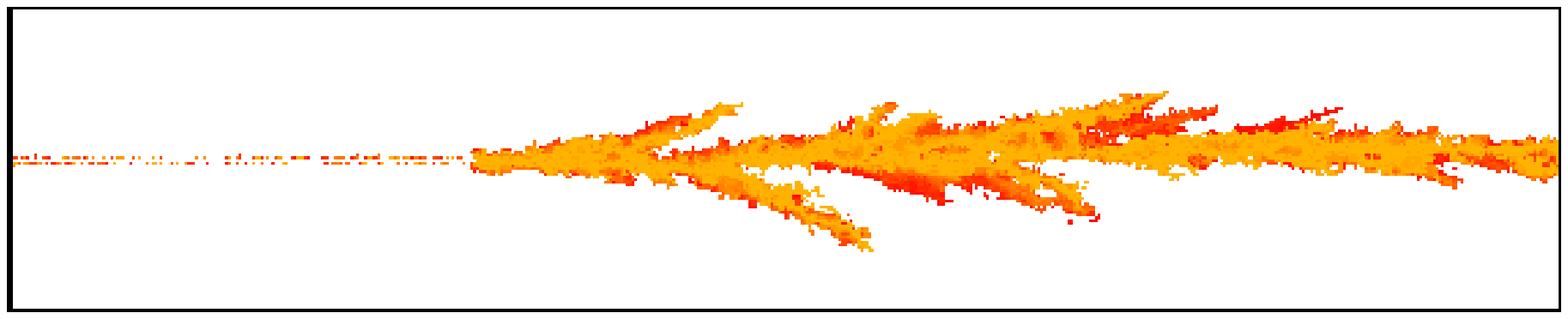, width=7cm}
}
\medskip
\centerline{
\epsfig{figure=\dirB 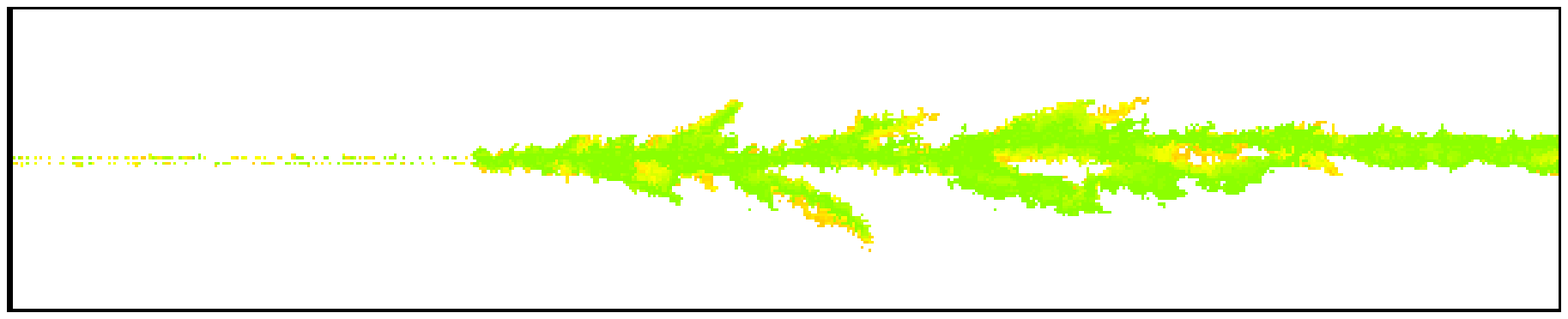, width=7cm}
}
\medskip
\centerline{
\epsfig{figure=\dirB 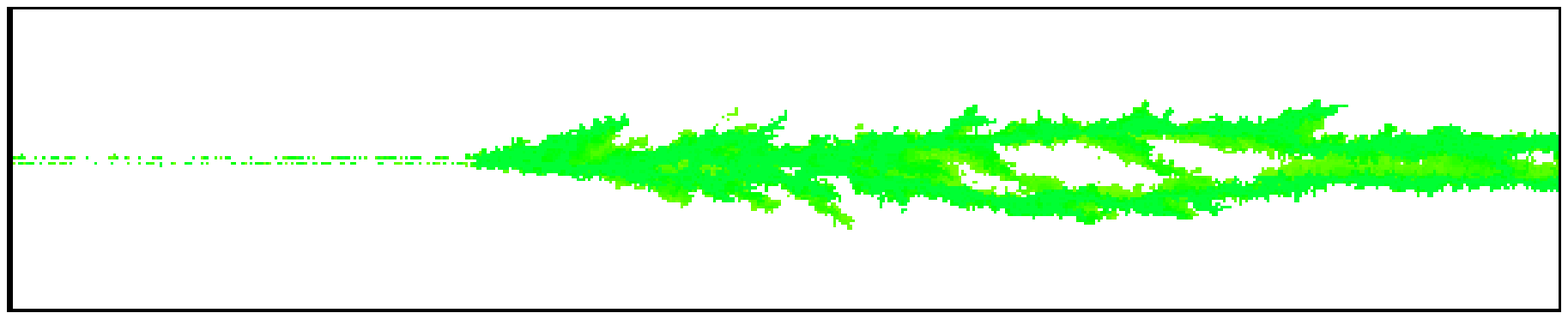, width=7cm}
}
\medskip
\centerline{
\epsfig{figure=\dirB 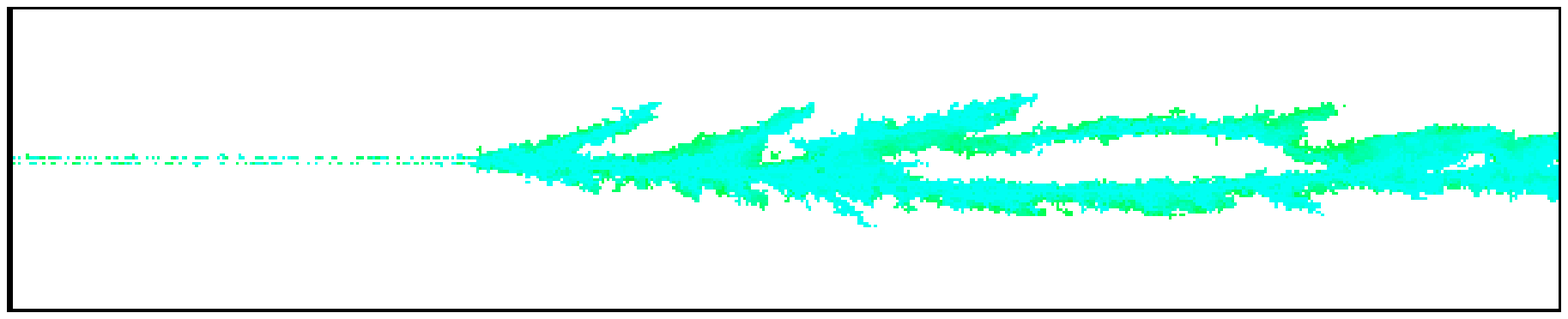, width=7cm}
}
\medskip
\centerline{
\epsfig{figure=\dirB 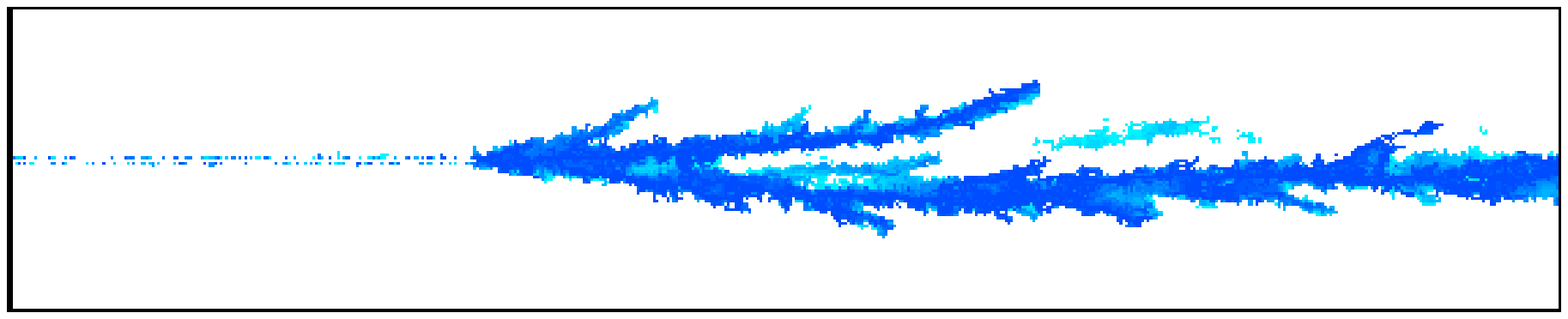, width=7cm}
}
\medskip
\centerline{
\epsfig{figure=\dirB 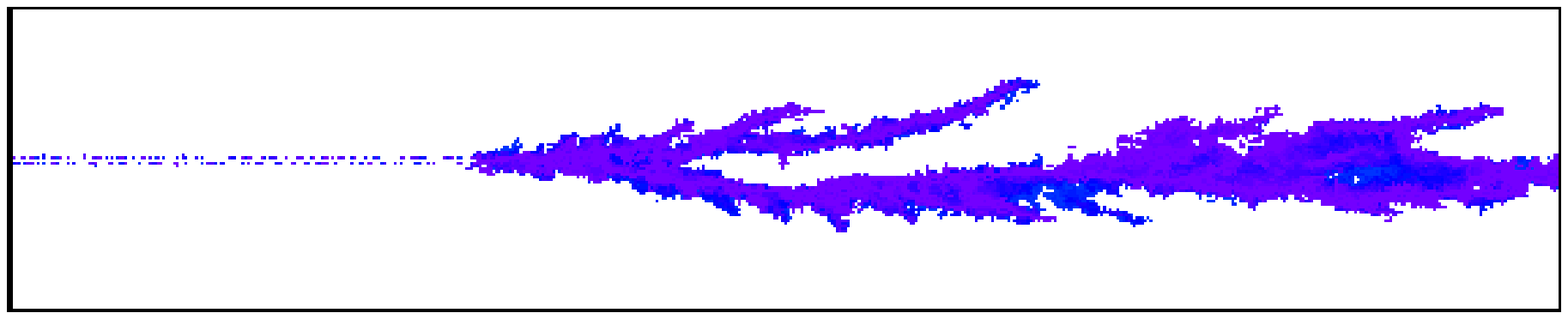, width=7cm}
}
\medskip
\centerline{
\epsfig{figure=\dirB 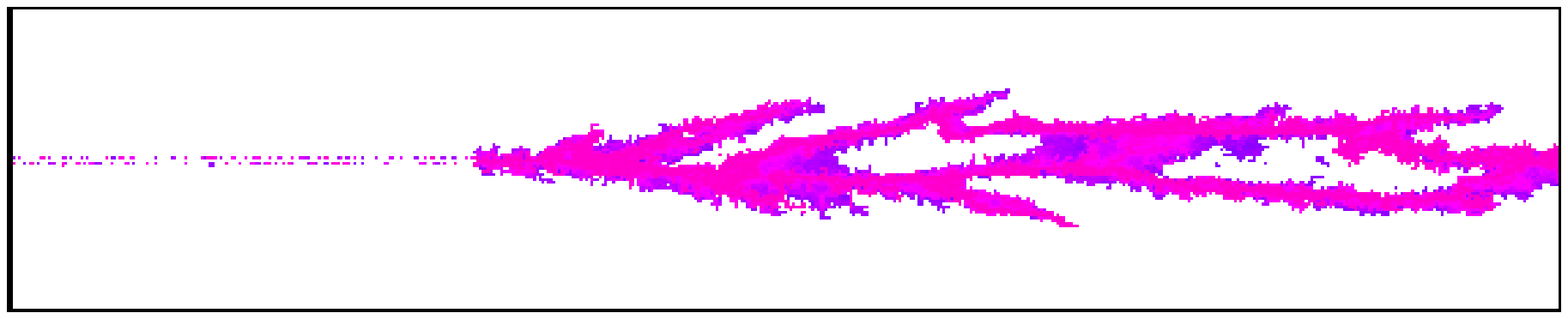, width=7cm}
}
\medskip
\centerline{
\epsfig{figure=\dirB 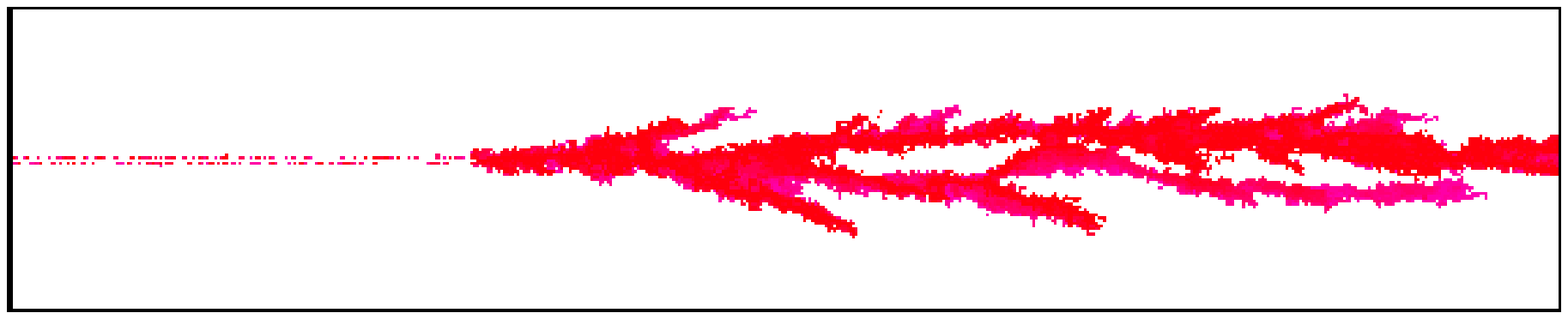, width=7cm}
}
\caption{(Color online) Slices of the fracture of figure 
\protect{\ref{fig:sequence}} 
viewed from the side.  Each slide corresponds to $1/8$ of the 
sample.  Shadings (and colors online) refer to the 
depth in the third dimension.}
\label{fig:sideview}
\end{figure}

\subsection{Measurement and selection of Crack Speed}

A first series of simulations was built using samples with sides $300\times60\times60$
tetrahedra, with a starting notch $100$ tetrahedra long. In these
simulations, the condition of fixed crack speed was released and the
Griffith criterion was used, but the condition of planarity was maintained
so that comparison could be made with the efficiency description.
A sample of three simulations were performed for each value of $\epsilon$:
above a threshold value of $0.7$, all simulations led to a completely
broken sample. Below this value, we found that cracks did not reach
the end of the sample.

The constraint of planarity was lifted in the second and third series
of simulations. The second had parameters matched to those of the
first, except that larger samples of 10 simulations were used to counter
the greater variability in results. The third series had larger size
simulations of $500\times120\times120$ tetrahedra and due to computational
cost was limited to samples of three simulations.

Crack speeds were measured with two different methods. In the \emph{histograms
method} a set of histograms was build, reporting the number of broken
tetrahedra at each distance from the crack notch (see figure \ref{fig:crackspeed-histo}).
Then, the intercepts with about $1/3$ of the average value of broken
tetrahedra per unit of crack advance (dashed line in figure) were
taken as a measure of the crack front position. The value of $1/3$
was chosen because that is where the histograms are typically steepest.
Plotting the position against time, the crack speed was retrieved.

In the \emph{average crack tip position method}, the average position
of the newly-broken tetrahedra in each time interval considered was
retrieved as a function of time, and the crack speed was measured
from its slope.

In the first set of simulations, corresponding to planar cracks, 
the values of the crack speed found from both methods are fully
compatible with the results found in our previous work\cite{PB-02} as 
shown from table
\ref{tab:crack-speeds}(a) and figure \ref{fig:crackspeed-energy}.
The results found are interesting not only for their
agreement with the results from the efficiency description, but also because
the measured values of $v/v_{R}$ are compatible with the crack speeds
measured in anisotropic materials,\cite{F-71,HB-66,WK-94}
in particular with the value of $0.9\,\,v_{R}$ 
measured for the sample of polymethilmetacrylate prepared with a weak
interface.\cite{WK-94} That value was supposed to be the best proof that 
the limiting crack speed for cracks is the Rayleigh speed. 
The microscopic structure
of polymethilmetacrylate is completely different from that of our
model, however due to the fact that the drop of the efficiency at
high speeds is dominated by the resonance at the Rayleigh speed
(which we expect it to be a common feature to all materials), we 
suggest that the limiting crack speed for a non-branching crack is
the Rayleigh speed \emph{only} in the limit of infinite loading, or 
$\epsilon\rightarrow\infty$. 
For finite loadings, the value of the crack speed measured corresponds
to that given from the efficiency description.  

The picture is notably different when the condition of planarity is
released in the second and third series of simulations. The disorder
dominates over the lattice anisotropy, with particularly resistant
(or particularly weak) tetrahedra prompting the crack to deflect out
of plane. For low driving forces, the resulting cracks are {}``fat''
(figure \ref{fig:loweps}) and the crack speed is considerably lower
than expected from the efficiency description. Crack speeds measured
by both methods are reported in table \ref{tab:crack-speeds}(b).
Low values of $\epsilon$ give a crack speed which is lower than expected
from the efficiency description, but compatible with some of the measurements
obtained both in simulations and in experiments as explained in section
\ref{sec:introduction}.
For higher values of $\epsilon$, the results of the two types of
measurements for the crack speed diverge, which we attribute to the
influence of crack branching. The average of the crack tip position
is an average of the new broken tetrahedra, which includes tetrahedra
which break along branches behind the crack front. As a result, the
average is affected by a systematic error which lowers the value from
that of the crack front. Where the speed measurements differ 
significantly, we take the histogram method to give the true speed of 
the crack front and the divergence of the methods to be an indicator 
of crack branching.

For the highest values of $\epsilon$ in the set of smaller simulations, 
the histogram method gives a value of the crack speed above that predicted from 
the efficiency function and, for $\epsilon=5$, above the Rayleigh
speed. This unexpected result is a consequence of the high level of
damage of the sample due to the high level of the driving force. Direct
inspection (see fig.~\ref{fig:sample-eps5}) shows that the sample
is broken, with branches reaching the sample's top and
bottom boundaries. The anomaly disappears upon increasing the size
of the sample. Simulations for samples $500\times120\times120$ tetrahedra
wide show, for low values of $\epsilon$, results compatible with
those of the smaller samples and, for all values of $\epsilon$, crack
speeds lower than the values given by the efficiency function (see
table \ref{tab:crack-speeds}(c)). Direct inspection (see fig.~\ref{fig:sample-eps5})
shows that although the damage is still heavy, in this case branches
no longer reach the sample boundaries. 

How do these results compare with those of other simulations and experiments?
Experiments are usually performed with carefully controlled loading.
The load is slowly increased and put just above the level beyond which
the macroscopic fracture develops. This corresponds to a low level
of $\epsilon$. This can help explain the compatibility of these
values of the crack speed with those measured in experiments. With
respect to the results of the planar case however, crack speeds
are lower. Although no major branches develop as testified by the
agreement in the values of the crack speed for the two type of measurement,
such drop in crack speed with respect to the efficiency prediction
must be connected with the mechanism of attempted branching which
is responsible for the {}``fat'' appearence of these cracks. For
$\epsilon=1$, the average width of a crack is $2.35$ tetrahedra,
indicating that more energy is needed for the crack to advance than
expected from the case of a the fully planar crack.

Higher values of $\epsilon$ do not correspond to the experimental
set up for the measurement of crack speeds. For such values, the crack
speed approaches the values given by the efficiency description. The
growth of the crack speed for increasing loading is similar to that
measured in other simulations \cite{PGLGS-00,FPGGL-02}. The high driving
force regimes are those which give rise to a non-zero roughness exponent
as we will see below.

\subsection{The shape of advancing cracks}

Figure \ref{fig:sequence} shows a sequence of frames of an advancing
crack for $\epsilon=3$, up to its final state, when the sample is
fully broken. Due to the magnitude of $\epsilon$, a considerable
level of branching is visible. Straight cracks are obtained for values
of $\epsilon\leq1$ (see later fig.~\ref{fig:sequence-eps}).

At first sight, a reasonable description for the phenomenon is that
of a main crack from which minor branches develop during the dynamics.
This is not a complete picture. Side views of the same final fracture
(see fig.~\ref{fig:sideview}) reveal that the crack is made by a
set of connected branches. The crack tip splits into two or more branches
which try to avoid each other, and force the crack to advance in a
non linear fashion. The particular fracture shown is also characterized
by two main branches running along much of the sample, and comparing
the different panels of figure \ref{fig:sideview} it can be seen
that both of these are part of the final fracture surface. Most branches
die out as they head towards the sample's boundaries. Those branches
that mantain their distance from the sample's boundaries manage to
travel through the sample building up its backbone. The development
of branches and then of the backbone, is clearly controlled by the
sample's boundaries which act as a guide to the branching process
and lead the whole crack in the forward direction.

%
\section{Roughness of fracture surfaces\label{sec:crack_roughness}}

\begin{figure}
\centerline{
\epsfig{figure=\dirB 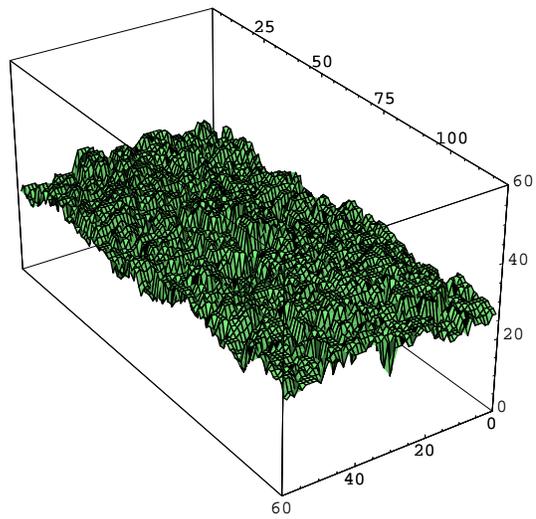, width=7cm}
}
\caption{Final fracture surface corresponding to a driving energy
$\epsilon = 3$.}
\label{fig:typ_surface}
\end{figure}

An example of a final fracture surface is shown in figure \ref{fig:typ_surface}.
The surface does not appear flat, and its roughness can be quantified
through the roughness exponent. For two points separated along the
direction of (global) crack proagation we have 
\[
\left\langle \left(h(x)-h(x')\right)^{2}\right\rangle \propto\left|x-x'\right|^{2\zeta_{x}}
\]
and similarly we define $\zeta_{z}$ for the scaling of height fluctuations
along the direction of the (global) crack edge. These scaling laws
can equivalently be probed by spatial power spectra, so that for a
cut along the $x$-direction we expect 
\[
\left\langle \widetilde{h}(k_{x})^{2}\right\rangle \propto\left|k_{x}\right|^{-1-2\zeta_{x}}.
\]
The scaling is expected to apply from local lengthscales up to of
order the (smallest) dimension of the sample. 

\begin{figure}
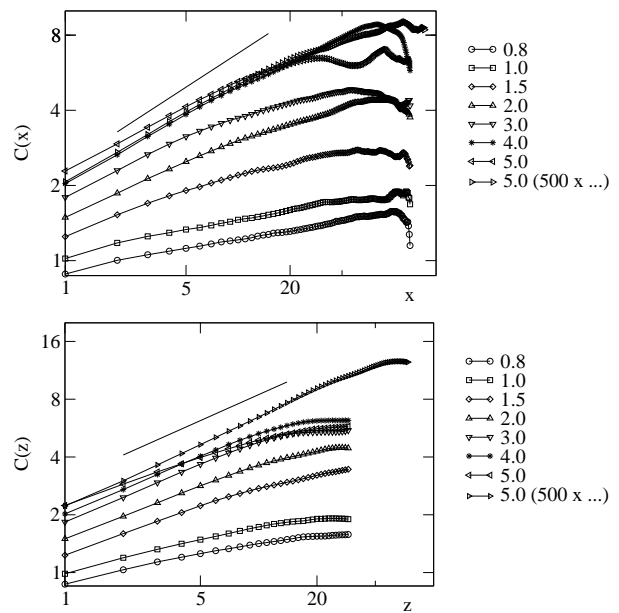

\centerline{
\epsfig{figure=\dirB RoughnessPlot-x.eps, width=8cm}
}
\medskip
\centerline{
\epsfig{figure=\dirB RoughnessPlot-z.eps, width=8cm}
}
\caption{Log-log plots of the height-height correlation functions along the
$\hat{x}$ direction (on the top), and the $\hat{z}$ direction (on the bottom) 
for the $300\times60\times60$ samples. For comparison,
also one of the correlation functions for the $500\times120\times120$
samples is shown. The above slope corresponds to roughness exponent $\zeta=0.45$.}
\label{fig:roughPlot}
\end{figure}

\begin{figure}
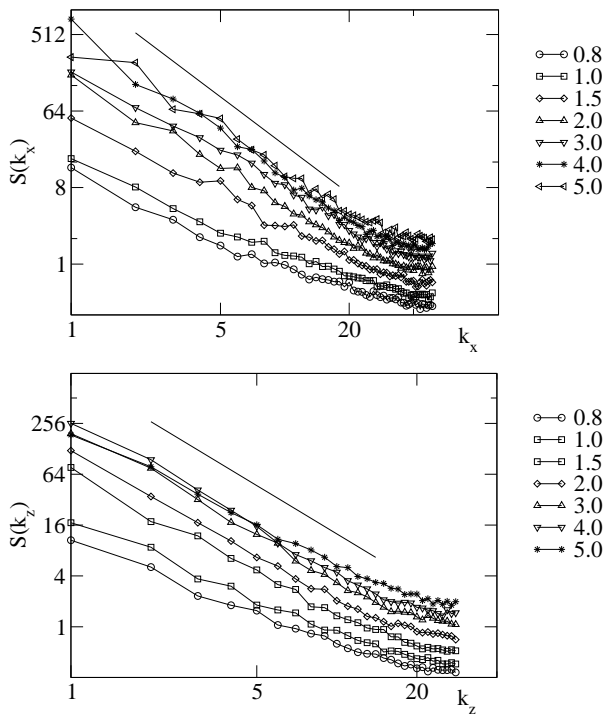

\centerline{
\epsfig{figure=\dirB RoughnessPlot-Fx.eps, width=8cm}
}
\medskip
\centerline{
\epsfig{figure=\dirB RoughnessPlot-Fz.eps, width=8cm}
}
\caption{Log-log plots of the height power spectra for cuts along the $\hat{x}$
direction (on the top), and the $\hat{z}$ direction (on the bottom)
for the $300\times60\times60$ samples. The above slope corresponds
to $\zeta=0.45$.}
\label{fig:roughPlotF}
\end{figure}

\begin{figure}
\centerline{
\epsfig{figure=\dirB 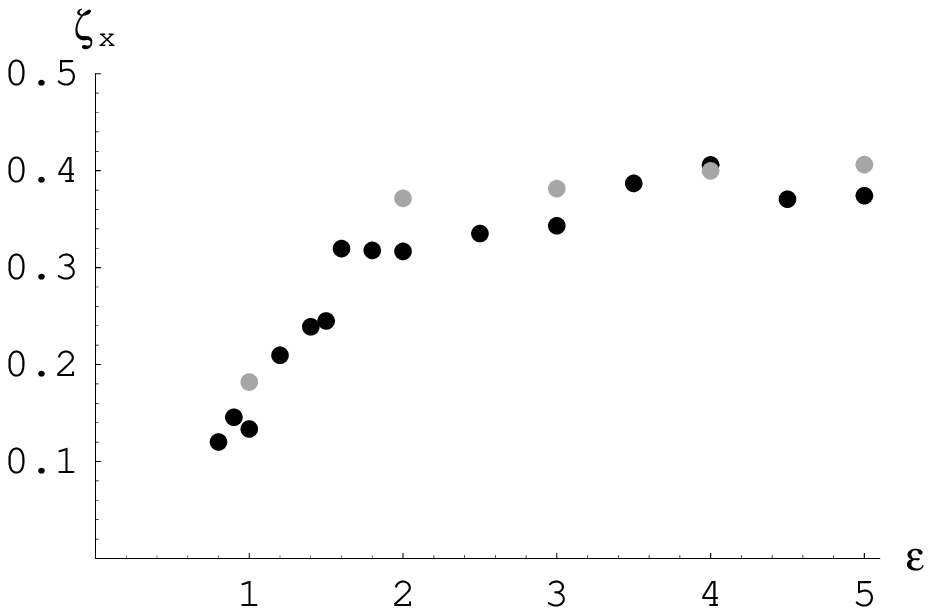, width=8cm}
}
\centerline{
\epsfig{figure=\dirB 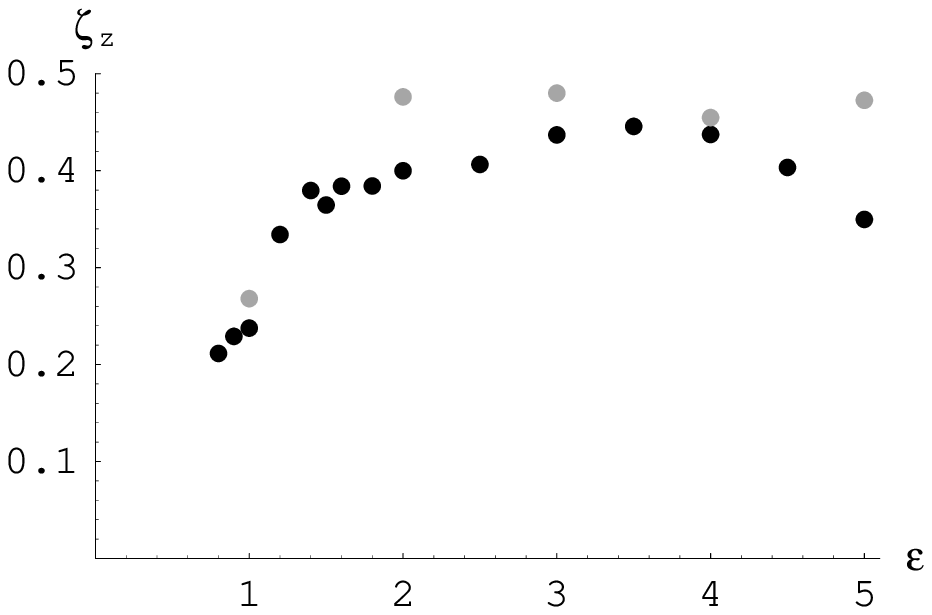, width=8cm}
}
\caption{Change
of the roughness exponent from the correlation functions, with the
driving factor $\epsilon$ for cuts along the $\hat{x}$ direction
(on the top) and the $\hat{z}$ direction (on the bottom). Results
for the $300\times60\times60$ samples are in black, those for the
$500\times120\times120$ samples are in light gray. \label{fig:rough-eps} }
\end{figure}

\begin{figure}
\centerline{
\epsfig{figure=\dirB 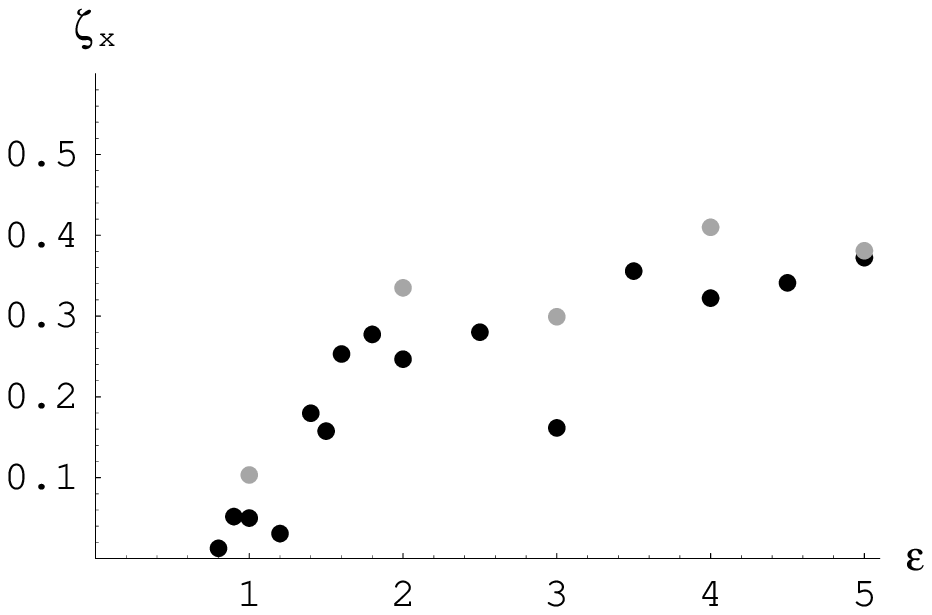, width=8cm}
}
\medskip
\centerline{
\epsfig{figure=\dirB 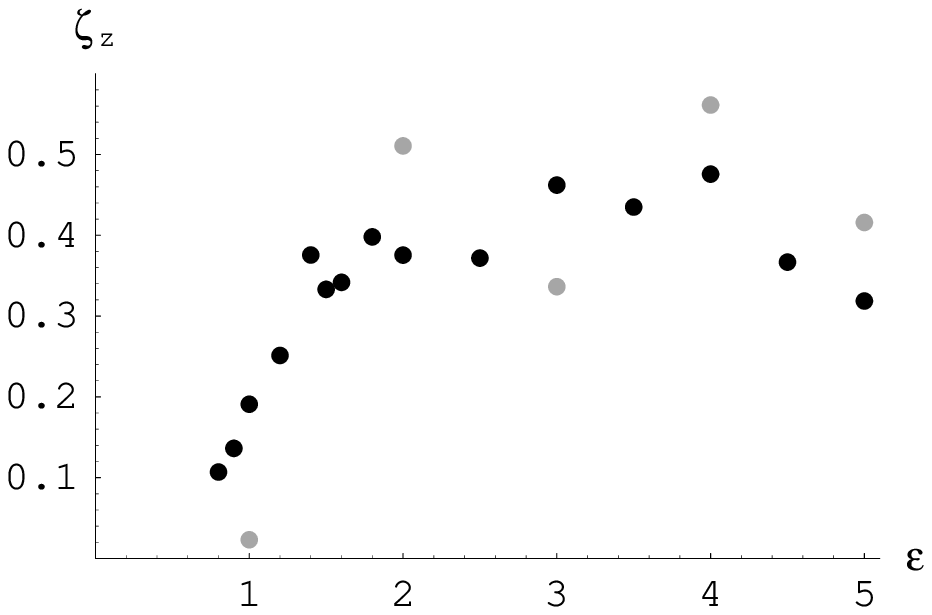, width=8cm}
}
\caption{Change of the roughness exponent from the power spectra, with the
driving factor $\epsilon$ for cuts along the $\hat{x}$ direction
(on the top) and the $\hat{z}$ direction (on the bottom). Results
for the $300\times60\times60$ samples are in black, those for the
$500\times120\times120$ samples are in light gray.}
\label{fig:roughF-eps}
\end{figure}

We find the measured roughness exponent varies systematically with the strength
of crack driving $\epsilon$, whilst there is relatively little difference
between measurements of the exponent in different directions or by
different methods. To limit the influence of the boundaries and of
the starting notch, we analysed a region limited to the central $80\times60$
tetrahedra for the $300\times60\times60$ samples, and $120\times120$
tetrahedra for the $500\times120\times120$ samples, equidistant
between the final boundary and the end of the starting notch.

Correlation functions for the $300\times60\times60$ samples, shown
in figure \ref{fig:roughPlot}, appear ordered in $\epsilon$ for
their slope. For increasing values of the driving energy, a region
of constant slope close to the origin develops in the corresponding
correlation function. The extension of this region is larger, the
higher the value of $\epsilon$. From its slope we have measured the
roughness exponent corresponding to each value of the driving force.
The roughness of the surface grows up to a limiting value, as shown
by the common slope of the highest curves in both figures. The loss
of slope of the highest curve corresponding to $\epsilon=5$ 
can be explained due to the approach of branches to the sample's boundaries
as described in the previous section.

Figure \ref{fig:roughPlotF} shows the power spectra of the same
cuts on the surface. Although the common roughness exponent is retrieved,
the characteristics of how this common slope builds up are less clear,
and data are more scattered.

Results for the roughness exponent for both sets of samples are reported
in figures \ref{fig:rough-eps} and
\ref{fig:roughF-eps}. Results from the power spectra are compatible
with the real space ones, although again more scattered. These
results show that the roughness exponent increases quite sharply from a value
close to zero, which we will see below can be connected with a
logarithmic scaling, up to a value between $0.4$ and $0.5$ for both
the $\hat{x}$ and $\hat{z}$ directions. Comparison between the values
of the roughness exponent along the two space directions shows a slight
difference which might be attributable to boundary effects, although data are insufficient 
to draw a definite conclusion.

\begin{figure}
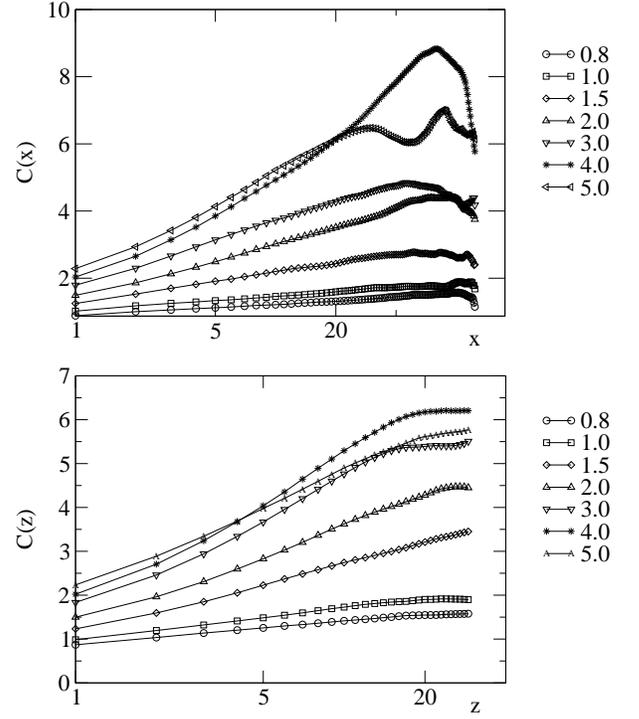

\centerline{
\epsfig{figure=\dirB RoughnessPlotLog-x.eps, width=8cm}
}
\medskip
\centerline{
\epsfig{figure=\dirB RoughnessPlotLog-z.eps, width=8cm}
}
\caption{Linear-log plots of the height-height correlation functions along
the $\hat{x}$ direction (on the top), and the $\hat{z}$ direction
(on the bottom) for the $300\times60\times60$ samples.
In both plots, the lowest curves corresponding to the lowest $\epsilon$ 
are straight, corresponding to a logarithmic scaling.}
\label{fig:roughPlot-log}
\end{figure}

For the lowest values of $\epsilon$ at which we could propagate cracks,
our roughness data are better described by a logarithmic roughness
law as shown by figure \ref{fig:roughPlot-log}. This is interesting
because it matches calculations by Ball and Larralde\cite{BL-95}
and subsequent calculations by Ramanathan, Erta\c{s} and Fisher\cite{REF-97} 
based on continuum elastic fracture mechanics for cracks at the threshold 
of propagation, as well as supporting experiments reported in 
Ref.~\onlinecite{LB-95}.  In continuum elastic fracture
mechanics these cracks are quasi-static, 
which we know from the efficiency description
is not the case for a structured material, whilst in the supporting
experiments\cite{LB-95} the overall propagation was kept slow
but locally crack acceleration could (and did) occur. Thus it appears
that the logarithmic law may apply rather generally to cracks propagating
marginally in three dimensions, without restriction to zero speed.

\begin{figure}
\centerline{
\epsfig{figure=\dirB 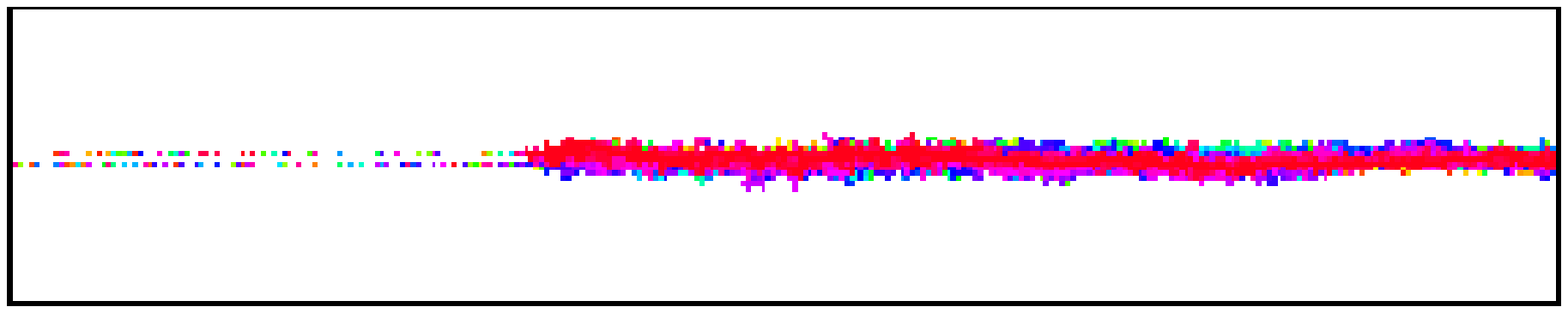, width=7cm}
}
\medskip
\centerline{
\epsfig{figure=\dirB 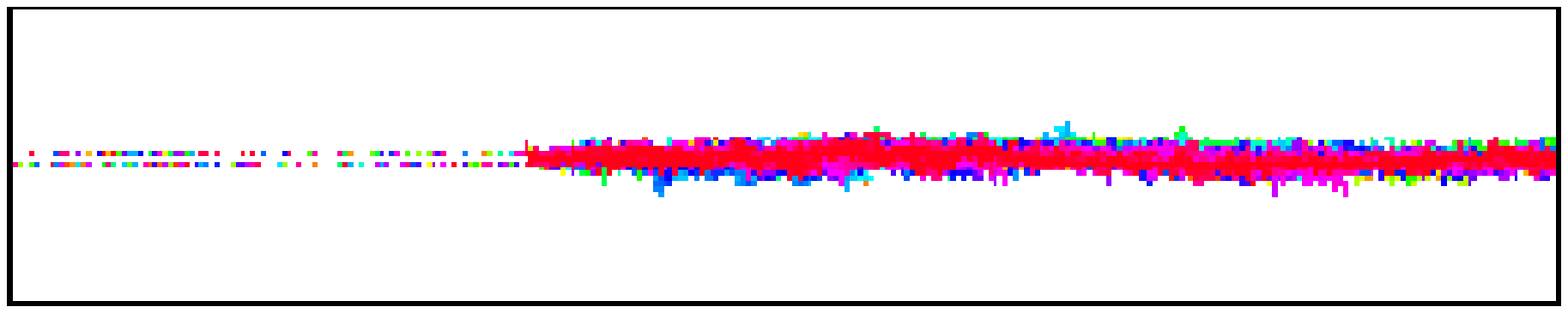, width=7cm}
}
\medskip
\centerline{
\epsfig{figure=\dirB 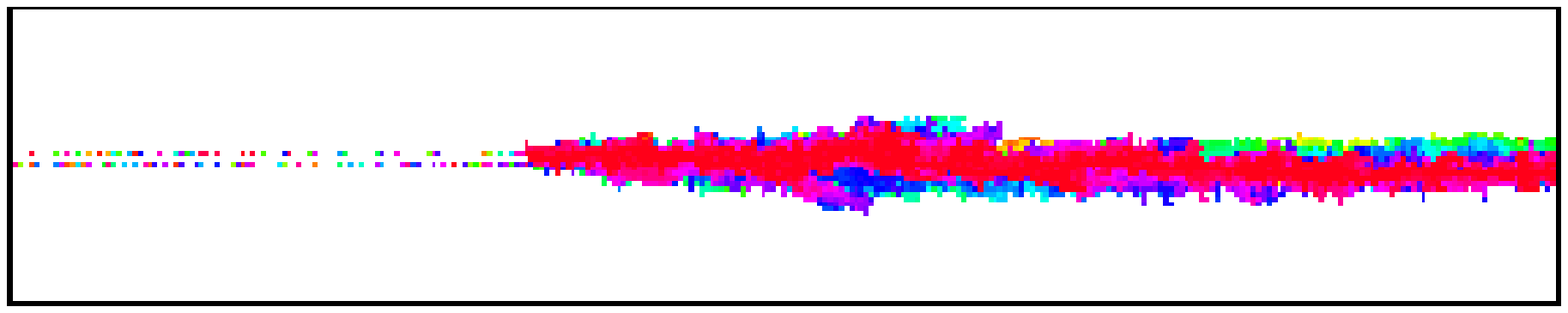, width=7cm}
}
\medskip
\centerline{
\epsfig{figure=\dirB 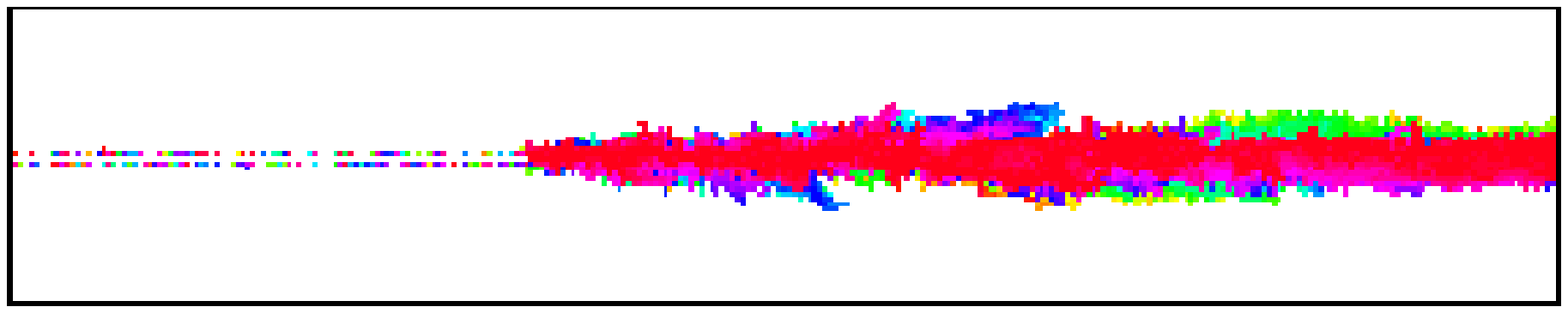, width=7cm}
}
\medskip
\centerline{
\epsfig{figure=\dirB 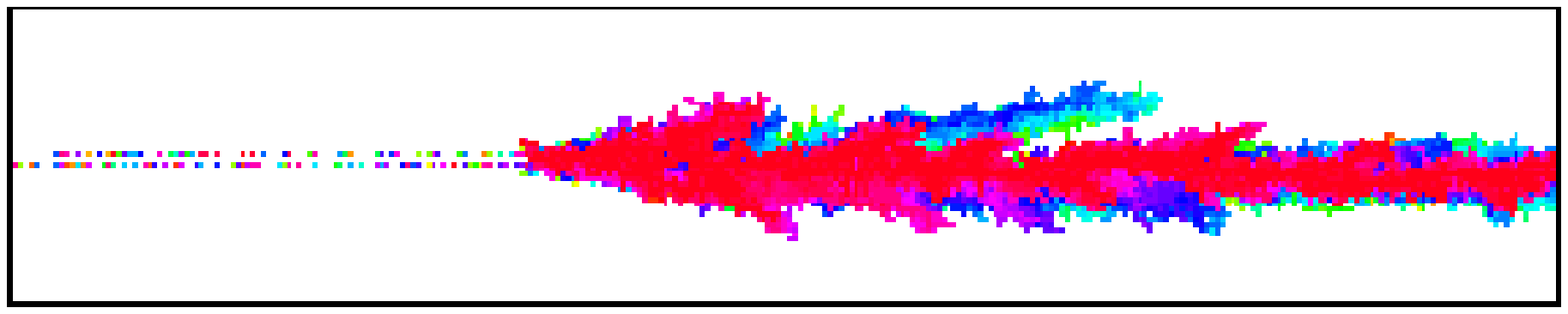, width=7cm}
}
\medskip
\centerline{
\epsfig{figure=\dirB 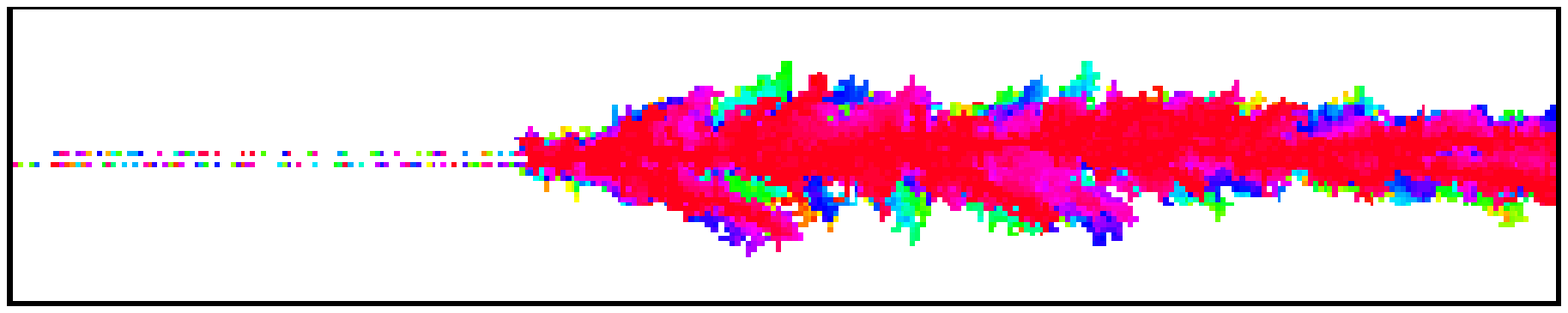, width=7cm}
}
\medskip
\centerline{
\epsfig{figure=\dirB 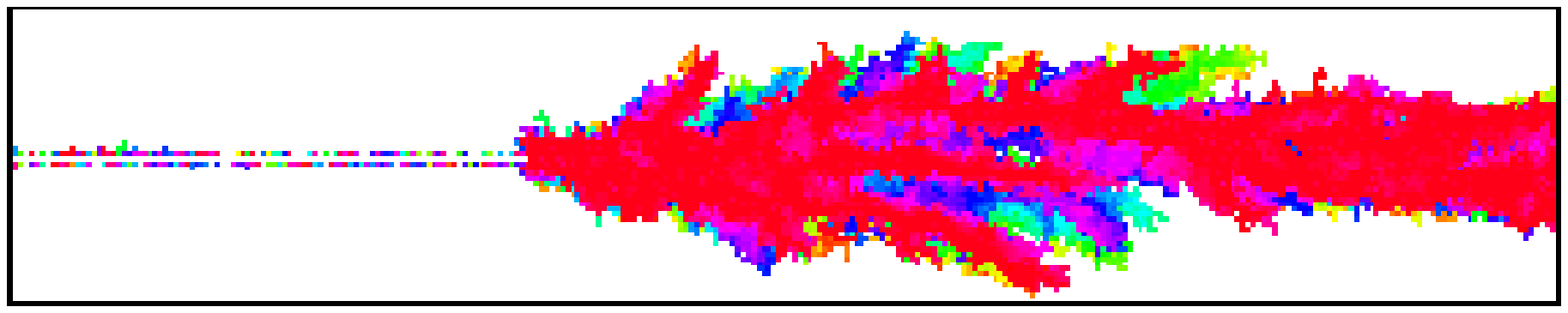, width=7cm}
}
\medskip
\centerline{
\epsfig{figure=\dirB sample-150x30x30-050.ps, width=7cm}
}
\caption{Shape of cracks in a sample $300\times60\times60$ tetrahedra wide 
(projections) for increasing values of $\epsilon$. Values
shown are from the top: $\epsilon=0.8$, $1.0$, $1.4$, $1.6$, $2.0$,
$3.0$, $4.0$, $5.0$.  Shading refers to the depth in the third dimension.}
\label{fig:sequence-eps}
\end{figure}

The increase in the fracture roughness can be visually connected to
the branching process by comparing these results to the sequence of
figure \ref{fig:sequence-eps}. The sequence shows that cracks are
flat for low values of $\epsilon$ within the limits of what disorder
allows. Macroscopic branches appear at a value of $\epsilon$ around
$1.4-1.6$ which corresponds to the increase of the roughness exponent
towards its limit value. The result should be compared with what observed
by Sharon, Gross and Fineberg\cite{SGF-95} in their experiment on the
branching instability. The shape of branches appeared to be compatible
with a power law shape with an exponent of about $0.7$. Our results
suggest that the appearence of branching increases the roughness 
to a value which in our case is between $0.4$ and $0.5$,
which corresponds to the one measured at short length scales.

\section{Simulation of disconnected fractures\label{sec:disconnected}}

\begin{figure}[b]
\centerline{
\epsfig{figure=\dirB 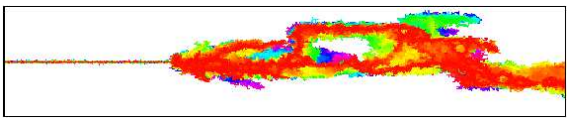, width=7.5cm}}
\caption{\label{fig:notconnected-crack}Sideview of the final fracture 
in the case of non-connected fracture as discussed in 
Section \protect\ref{sec:disconnected}, for a $500\times120\times120$ 
tetrahedra wide sample and $\epsilon=5$.
Shading refers to the depth in the third dimension.}
\end{figure}

\begin{table*}
\begin{ruledtabular}
\begin{tabular}{c c c c c}
  & \multicolumn{2}{c}{Height-height correlation function} & \multicolumn{2}{c}{Power spectrum} \\
\hline
$\epsilon$ &  $\zeta_x$ &  $\zeta_z$  &  $\zeta_x$ &  $\zeta_z$  \\
\hline
 $1.0$ & $0.153 \pm 0.008$ & $0.265 \pm 0.006$ & $0.058 \pm 0.033$ & $0.030 \pm 0.060$ \\
 $2.0$ & $0.330 \pm 0.002$ & $0.419 \pm 0.003$ & $0.280 \pm 0.050$ & $0.402 \pm 0.087$ \\
 $3.0$ & $0.339 \pm 0.003$ & $0.445 \pm 0.005$ & $0.341 \pm 0.042$ & $0.661 \pm 0.050$ \\
 $4.0$ & $0.473 \pm 0.005$ & $0.387 \pm 0.001$ & $0.525 \pm 0.011$ & $0.398 \pm 0.054$ \\
 $5.0$ & $0.418 \pm 0.002$ & $0.452 \pm 0.004$ & $0.509 \pm 0.043$ & $0.526 \pm 0.094$ \\
\end{tabular}
\end{ruledtabular}
\caption{\label{tab:notconnected-rough-eps}Roughness
exponent for different values of $\epsilon$ in the case of non-connected
fractures, as discussed in Section \protect\ref{sec:disconnected}. 
All these results have been obtained from simulations of samples
$500\times120\times120$ tetrahedra wide. The second and third column
correspond to the roughness exponent measured from the height-height
correlation functions along cuts in the $\hat{x}$ and the $\hat{z}$
directions. The fourth and fifth columns correspond to the roughness
exponent measured from the power spectra.}
\end{table*}

The limit value for the roughness exponent at about $0.45$ corresponds
to the one measured in molecular dynamics simulations by Nakano, Kalia 
and Vashishta\cite{NKV-95} 
in microcrack advance, where they suggest microcrack coalescence 
as the mechanism leading to higher values $\sim0.75$.\cite{NKV-95} 
As all our simulations
above have been performed imposing the connection of the advancing
crack, which prevents the formation of multiple cracks within the
sample, we could expect to have the same coexistence of the two values
for the roughness exponent if we release this additional condition.
Its removal allows the creation of diffuse damage (commonly referred
to as {}``dust''), composed of isolate broken tetrahedra scattered
throughout the sample. The level of such dust increases with $\epsilon$,
as the number of breakable tetrahedra increases with $\epsilon$.
In fig.~\ref{fig:notconnected-crack} the case for $\epsilon=5$
is shown. The outcome has been cleaned of the dust which, due to the
extreme high value of $\epsilon$, sums up to a number of broken bonds
equivalent to those shown, but scattered through the whole sample.
The final fracture is characterized by large jumps due to microcracks
which have connected during the dynamics, suggesting a similar interpretation
to that of Nakano \emph{et al.} Measurements of the corresponding
roughness exponent for the $500\times120\times120$ samples however,
reported in table \ref{tab:notconnected-rough-eps}, do not show any
relevant difference from those of connected fractures. This may just reflect
the length scale at which microcrack coalescence develops in our
simulations being larger than the length scale over which our roughness measurements
are performed. To check any increase in the roughness exponent
associated with the lengthscale of crack coalescence evident in fig.~\ref{fig:notconnected-crack} an increase in the system size would certainly be
necessary.

\section{Conclusions\label{sec:conclusions}}

The release of the condition of fixed crack speed has been the first
test for the efficiency description of the selection of the crack
speed. The efficiency description shows, in contrast to the theoretical 
description
of the continuum theory, that the Rayleigh speed is the ultimate crack
speed only in presence of an infinite loading. For finite loading
applied, the crack speed selected follows the efficiency description
in presence of consistent anisotropy or lattice trapping.
That the terminal crack speed would be the Rayleigh speed only
for an infinite loading had already been suggested by 
Xu and Needleman.\cite{XN-94}
However, here this result comes out naturally from the shape of the 
efficiency, due to the Griffith criterion (\ref{eq:Griffith-discr}).

When the fracture geometry is released from planar, the terminal crack speed in our simulatiosn is much lower
than expected from the efficiency description, and corresponds to the range of crack speed measured in experiments. 
The analysis of the crack shape has shown that even at the lowest loadings,
cracks attempt to branch and this tends to thicken the crack. Hence,
the amount of energy which is delivered into fracture work per unit
of crack advance has to increase, and the energy radiated has to decrease.
This inevitably leads to a change in the maximum speed achieved. Although
the basic idea is clear, more work is required to relate
the effective amount of energy radiated with the branching mechanism
and hence with a criterion of speed selection for non-planar cracks.

When the energy is sufficient, branching becomes a macroscopic phenomenon
and appears to be the basic mechanism through which the roughness
exponent of a crack surface builds up. The dynamics of the macroscopic
branches determine the final fracture shape: branches build the backbone
of the whole crack. During the crack advance branches try to avoid
each other whilst the boundaries of the sample lead them into the
forward direction, creating a preferential direction for their advance. 
All the branches that greatly deviate from this direction
die out and do not belong to the final fracture surface.

When the driving energy is low, the self-affine properties of the
crack surface are also compatible with the scenario of logarithmic roughness
suggested in Ref.~\onlinecite{LB-95,BL-95} and \onlinecite{REF-97}. 
The condition of quasi-static cracking is explicit in the latter 
in terms of the fracture energy supplied to the crack being barely 
in excess of the fracture toughness. This
however does not correspond simply to having a crack with little kinetic
energy. On the contrary, because of the efficiency description, low
values of $\epsilon$ correspond to cracks which barely have energy
to advance, but which travel at high speed. This could clarify how
a logarithmic roughness predicted for a quasi-static crack can be found 
in some of our simulations of dynamic cracks.

The roughness exponent grows rapidly with the driving energy $\epsilon$,
towards a maximum of $\sim0.45$ corresponding to that measured in
experiments on short length scales.  
Recent experiments\cite{LS-98,MSLV-98} have shown that fracture surfaces
could be affected by anomalous scaling.  This in turn has been 
interpreted\cite{MBSV-02} as the mark of the anisotropy of the roughness 
exponent measured at large lenght scales.
Our simulations show a slight anisotropy of the final fracture surface,
which might be an indication of similar behaviour on short length scales, 
and calls for further attention in both simulations and experiments.

We have not been able to corroborate the suggestion of 
Ref.~\onlinecite{NKV-95} that the higher roughness exponent for large 
scales results from microcrack coalescence, but this may be due 
to the limited range of lengthscales we have been able to access in 
these simulations.  We estimate we would require to gain one to two 
orders of magnitude in distance range to resolve this issue with the present model.

\begin{acknowledgments}
We acknowledge financial support of EU contract No. ERBFMRXCT980183.  
\end{acknowledgments}



\end{document}